\pdfoutput=1

\documentclass[final,10pt]{elsarticle}

\usepackage[utf8]{inputenc}
\usepackage{xcolor}
\usepackage{amsthm}
\usepackage{amsfonts}
\usepackage{amsmath,amssymb,mathtools}
\usepackage{helvet}
\usepackage[font=scriptsize]{subcaption}
\usepackage{fullpage}
\usepackage{array}

\newcommand{\mv}[1]{{\boldsymbol{#1}}}
\newcommand{\dmv}[1]{{\dot{\boldsymbol{#1}}}}
\newcommand{\hmv}[1]{{\hat{\boldsymbol{#1}}}}
\newcommand{\tmv}[1]{{\tilde{\boldsymbol{#1}}}}
\newcommand{\bmv}[1]{{\bar{\boldsymbol{#1}}}}
\newcommand{\norm}[1]{\left\lVert #1 \right\rVert}
\newcommand{\minimize}[1]{\underset{#1}{\mathrm{minimize \ \ }}}
\newcommand{\sgn}{\mathop{\mathrm{sgn}}}

\def\xref {\boldsymbol{x}_\mathrm{ref}}
\def\xhat {\hat{\boldsymbol{x}}}
\def\RR {\mathbb{R}}

\makeatletter
\def\ps@pprintTitle{%
 \let\@oddhead\@empty
 \let\@evenhead\@empty
 \def\@oddfoot{}%
 \let\@evenfoot\@oddfoot}
\makeatother

\newtheorem{theorem}{Theorem}[section]
\newtheorem{remark}[theorem]{Remark}

\begin{document}
\begin{frontmatter}

\title{Preserving general physical properties in model reduction of dynamical systems via constrained-optimization projection}

\author[tum]{Alexander Schein\corref{tumcor}}
\ead{schein@mhpc.mw.tum.de}
\author[uw,sandia]{Kevin T. Carlberg\corref{sandiacor}}
\ead{ktcarlb@uw.edu}
\author[nd]{Matthew J. Zahr\corref{ndcore}}
\ead{mzahr@nd.edu}

\address[tum]{Technical University of Munich}
\address[sandia]{Sandia National Laboratories}
\address[uw]{University of Washington}
\address[nd]{University of Notre Dame}

\cortext[tumcor]{Parkring 35, 85748 Garching b. M{\"u}nchen}
\cortext[sandiacor]{7011 East Ave, Livermore, CA 94550}
\cortext[ndcore]{Notre Dame, Indiana, USA 46556}


\begin{abstract}
Model-reduction techniques aim to reduce the computational
complexity of simulating dynamical systems by applying a (Petrov--)Galerkin
projection process that enforces the dynamics to evolve in a low-dimensional
subspace of the original state space. Frequently, the resulting reduced-order
model (ROM) violates intrinsic physical properties of the original full-order
model (FOM) (e.g., global conservation, Lagrangian structure, state-variable
bounds) because the projection process does not generally ensure preservation
of these properties. However, in many applications, ensuring the ROM preserves
such intrinsic properties can enable the ROM to retain physical meaning and 
lead to improved accuracy and stability properties.  In this work, we present
a general constrained-optimization formulation for projection-based model
reduction that can be used as a template to enforce the ROM to satisfy
specific properties on the kinematics and dynamics.
We introduce constrained-optimization formulations
at both the time-continuous (i.e., ODE) level, which leads to a constrained
Galerkin projection, and at the time-discrete level, which
leads to a least-squares Petrov--Galerkin (LSPG) projection, in the context of linear multistep
schemes. We demonstrate the ability of the proposed formulations to equip ROMs
with desired properties such as global energy conservation and bounds on the
total variation.
\end{abstract}

\end{frontmatter}

\section{Introduction}

Serving as a bridge to overcome the prohibitive computational cost of
simulating large-scale, parameterized dynamical systems in time-critical
scenarios, projection-based model reduction has gained popularity and yielded
successes across problems in many applications, including fluid dynamics,
structural dynamics, and reacting flows. In the context of proper orthogonal
decomposition (POD) \cite{Rathinam20031893} or reduced-basis method \cite{Quarteroni20111} approaches, a computationally costly \textit{offline
stage} is first executed in which the parameterized dynamical system, or
\textit{full-order model} (FOM), is simulated at multiple parameter instances.
Then, the state snapshots arising from these training simulations are employed
to construct a low-dimensional affine trial subspace of the original state
space; recent approaches have also investigated constructing low-dimension
nonlinear trial manifolds \cite{lee2018model} from these snapshots. A
parameterized dynamical system of lower dimension, called the reduced-order
model (ROM) is computed by performing a (Petrov--)Galerkin projection process
of the original dynamical system onto this subspace; as a result, the ROM's
solution is restricted to evolve in the low-dimensional trial subspace.  If
the FOM operators are nonlinear in the state or non-affine in the parameters,
then an additional `hyper-reduction' approximation is required to ensure the
resulting ROM simulation incurs an operation count that is independent of the
FOM state-space dimension. Computational-cost savings are achieved in the
\textit{online stage} by simulating the low-cost ROM at new parameter
instances.

The ROM approximation typically does not preserve important intrisic properties
of the dynamical system, e.g., global conservation and hyperbolicity in the
context of (semi-discretized) conservation laws \cite{Carlberg2018280} and
energy conservation in the context of solid dynamics \cite{an2008optimizing,Farhat2014625}. Failure of the ROM to
preserve these properties not only raises questions about the physical
interpretability of the resulting solution, it also has been attributed to
large state-space errors and lack of stability \cite{Carlberg2015B153}, which
has motivated research on structure-preserving model reduction for different
classes of dynamical systems. One class of such approaches derives governing
equations for the ROM in a manner that preserves the structure of the original
dynamical system, for example by preserving Lagrangian structure \cite{Carlberg2015B153,Lall2003304}, the principle of virtual work \cite{Farhat2014625,Farhat20151077}, \mbox{(Port-)}Hamilton structure \cite{Polyuga2010665,Afkham2017A2616,peng2016symplectic}, the principle of mass conservation
\cite{Lehrenfeld2019}, or positivity of contact forces in adhesion-free normal
contact \cite{balajewicz2016projection}. Researchers have pursued
several works on structure-preserving model reduction in computational fluid
dynamics. Refs.~\cite{sanderse2020non,afkham2020conservative} investigate conservative ROMs for fluid flows based on preservation of operator
skew-symmetry.  In
Ref.~\cite{mohebujjaman2019physically}, the authors enhance a previously
introduced strategy for projection-based ROM construction with data-driven
correction by physically motivated constraints ensuring energy dissipation by
the correction, while Ref.~\cite{loiseau2018constrained} proposes a data-driven regression model with an
energy-conservation constraint on the nonlinear operator of the Navier--Stokes
equations. 
Ref.~\cite{rowley2004model} proposes a technique to preserve the stability property of an equilibrium point at
the origin by applying a physically interpretable energy-based inner
product for construction of a compressible flow projection-based ROM.
Ref.~\cite{balajewicz2013low} constructs a
projection-based ROM incorporating a power balance constraint aiming at
physical representation of the incompressible flow energy cascade, and
Ref.~\cite{mohebujjaman2017energy} demonstrates energy conservation 
by a Stokes extension of boundary conditions.

Another strategy that is gaining increasing attention is the incorporation of equality and inequality
constraints that explicitly impose desired behavior in the kinematics and dynamics, such as
bounds on the coefficients of POD modes \cite{Fick2018214},
bounds on the temperature of a reacting flow \cite{Huang2019},
boundary-condition constraints \cite{Cao2019679}, conservation constraints for 
finite-volume models \cite{Carlberg2018280,lee2019deep},
constraints on the gas void fraction for bubbling fluidized beds \cite{Reddy201718}, and constraints on aerodynamic coefficients (lift, drag,
moment) for steady flows \cite{Zimmermann2014255}.

To our knowledge, all existing approaches for structure-preserving model
reduction focus on enforcing \textit{problem-specific properties} on the ROM solution;
there is no general model-reduction formulation that can be applied to enforce a
variety of structure-preserving constraints. As such,
researchers currently must re-derive a structure-preserving projection scheme
for each problem of interest. The goal of this work is to fill this gap by
developing a general projection-based model-reduction formulation
that can be employed to enforce a variety of general constraints; as such, it
may be used as a template to define a particular instance of a
structure-preserving ROM. Our approach comprises a novel
constrained-optimization-based formulation applicable to dynamical systems
expressed as systems of parameterized ordinary differential equations (ODEs).
The formulation supports the following:
	\begin{itemize}
		\item \textit{Constrained Galerkin and least-squares Petrov--Galerkin
			(LSPG) projection}. We propose two
			different constrained-optimization formulations: one that operates on
			the (time-continuous) parameterized system of ODEs (i.e., Galerkin
			projection), and one that operates on the (time-discrete) parameterized
			system (i.e., least-squares Petrov--Galerkin projection)
			\cite{carlberg2017galerkin}.
		\item \textit{Equality and inequality constraints}. Given that desired ROM
			properties might associate with both equality (e.g., conservation of
			energy) and
			inequality (e.g., diminishing total variation in time) constraints, the proposed formulation incorporates both of
			these types of constraints. 
	\item \textit{Constraints on both kinematics and dynamics}. Given that
		desired ROM properties might involve both \textit{dynamics constraints} that
			act on both the
			state and velocity (e.g., vanishing sum of ODE-residual entries) and \textit{kinematics constraints}
			that act exclusively on the 
			state (e.g., bounded state total variation), our formulation supports both types of
			constraints.
	\end{itemize}

Additionally, we employ this formulation to derive four different
problem-specific constraints and perform numerical experiments on a number of
parameterized dynamical systems that result from the semi-discretization of
partial differential equations: 1D Burgers' equation, 1D Euler equations, and
the 2D diffusion equation.  These numerical experiments demonstrate that
the proposed formulation can be used to derive structure-preserving ROMs that
can be equipped with the desired properties.

Because the focus of this paper is on developing a novel projection scheme for
model reduction, we do not directly consider hyper-reduction in this initial
work; the reader is referred to Ref.~\cite{Carlberg2018280}, which derived hyper-reduction
approaches for conservative reduced-order models that can be considered a
specific instance of the present framework.\footnote{Unsurprisingly, the key
idea is that any nonlinear terms that appear in either the objective function
or constraints must be approximated via hyper-reduction in order to realize
computational-cost reduction.} As such, we employ the general purpose optimization and root
finding functionality from the package \texttt{SciPy} \texttt{optimize} \cite{2020SciPy-NMeth} in the numerical results.

The remainder of this paper is organized as follows. Section
\ref{sec:Full-order model} formulates the parameterized dynamical system that
defines the FOM and introduces the class of linear multistep schemes we
consider
for temporal discretization. Section \ref{sec:Dimensional reduction} recalls
traditional Galerkin and LSPG projections and their associated (unconstrained)
minimization problems. Section \ref{sec:Constrained
optimization-based dimensional reduction} introduces the proposed
constrained-optimization formulations at both the time-continuous and
time-discrete levels. Section \ref{sec:Specific constraints} derives specific exemplary
constraints using this general framework. Section
\ref{sec:Numerical experiments} presents three numerical experiments that
employ these constraints.
Finally, Section \ref{sec:Conclusion} concludes the paper.

\section{Full-order model}
\label{sec:Full-order model}

We begin by introducing the FOM, which corresponds to a parameterized system of ordinary differential
equations (ODEs):
\begin{equation}
\label{eq:FOM}
\dmv{x} = \mv{f}(\mv{x}, t; \mv{\mu}), \qquad \mv{x}(0; \mv{\mu}) = \mv{x}^0(\mv{\mu}),
\end{equation}
where $t \in [0, T]$ denotes the time with $T \in \mathbb{R_+}$ denoting the
final time, $\mv{\mu} \in \mathcal{D}$ denotes a vector of parameters that
lives in a 
parameter domain $\mathcal{D} \subset \mathbb{R}^{n_{\mu}}$, $\mv{x}: [0, T]
\times \mathcal{D} \to \mathbb{R}^N$ denotes the time-dependent, parameterized
dynamical-system state (i.e., FOM ODE solution) implicitly defined as the solution to equation
\eqref{eq:FOM}, $\mv{x}^0: \mathcal{D} \to \mathbb{R}^N$ denotes the initial
condition, and $\mv{f}: \mathbb{R}^N \times [0, T] \times
\mathbb{R}^{n_{\mu}} \to \mathbb{R}^N$ denotes the velocity. As expressed in
equation \eqref{eq:FOM}, we typically drop the arguments
$(t; \mv{\mu})$ from the state $\mv{x}(t; \mv{\mu})$ and its time derivative
$\dmv{x}(t; \mv{\mu})$ for notational convenience. In residual form, we can
express equation \eqref{eq:FOM} as
\begin{equation}
\mv{r}(\dmv{x}, \mv{x}, t; \mv{\mu}) = \mv{0},
\label{eq:FOM in residual notation}
\end{equation}
where
\begin{equation}
\label{eq:time-continuous full-order residual}
\mv{r}: (\mv{v}, \mv{\xi}, \tau; \mv{\nu}) \mapsto \mv{v} - \mv{f}(\mv{\xi}, \tau; \mv{\nu})
\end{equation}
with $\mv{r}: \mathbb{R}^{N} \times \mathbb{R}^N \times [0, T] \times
\mathbb{R}^{n_{\mu}} \to \mathbb{R}^N$ denoting the time-continuous residual.

Computing a numerical solution to equation \eqref{eq:FOM} requires introducing a
time-discretization method. In this work, we consider the class of linear
multistep schemes \cite{carlberg2017galerkin}. Application of a linear
multistep scheme to \eqref{eq:FOM} yields the time-discrete dynamical system
(O$\Delta$E) corresponding to a sequence of algebraic equations 
\begin{equation}
\label{eq:time discrete residual}
	\mv{r}^{n}(\mv{x}^{n}; \mv{\mu}) = \mv{0},\quad n=1,\ldots,N_T,
\end{equation}
where 
\begin{equation}
\label{eq:linear multistep scheme}
\begin{split}
\mv{r}^{n}: (\mv{\xi}; \mv{\nu}) \mapsto  \alpha_0 \mv{\xi} &- \Delta t \beta_0 \mv{f}(\mv{\xi}, t^{n}; \mv{\nu})
+ \sum_{j=1}^{k} \alpha_j \mv{x}^{n-j} - \Delta t \sum_{j=1}^{k} \beta_j \mv{f}(\mv{x}^{n-j}, t^{n-j}; \mv{\nu}),\quad n=1,\ldots,N_T
\end{split}
\end{equation}
with $\mv{r}^{n}: \RR^{N} \times \RR^{n_{\mu}} \to \RR^N$,
$n=1,\ldots,N_T$ denoting the
time-discrete residual and $\Delta t=T/N_T$ denoting the times step
(assumed to be constant for notational simplicity). Selection of the coefficients $\alpha_j$,
$j=0,\ldots,k$ with
$\sum_{j=0}^k \alpha_j = 0$ and $\beta_j$,
$j=0,\ldots,k$ returns a
particular linear multistep scheme; the case $\beta_0 = 0$ yields an
explicit scheme. Solving equation \eqref{eq:time discrete residual} yields
the parameterized
time-discrete dynamical-system state
(i.e., FOM O$\Delta$E solution)
$\mv{x}^{n}(\mv{\mu}) \approx \mv{x}((n)\Delta t;
\mv{\mu})$, $n=1,\ldots,N_T$.

\section{Unconstrained dimensional reduction}
\label{sec:Dimensional reduction}

We now review traditional Galerkin and LSPG projection schemes and recall
their associated (unconstrained) minimization properties that we will build
upon to define the proposed formulation.

\subsection{Low-dimensional-subspace approximation}
\label{sec:Approximation on low-dimensional subspaces}

Given a low-dimensional subspace spanned by the columns of the
orthogonal reduced-order basis matrix 
$\mv{\Phi} \in  V_p(\RR^{N})$, where 
$V_m(\RR^{n})$ denotes the set of orthogonal $n\times m$ real-valued matrices
(the Stiefel manifold)
and $p\ll N$,
projection-based ROMs approximate the kinematics of the problem by restricting
the state to reside a low-dimensional affine subspace, which leads to the
time-continuous ROM
state and velocity 
\begin{align}
\tmv{x}(t, \mv{\mu}) &\coloneqq \mv{x}_{\mathrm{ref}}(\mv{\mu}) + \mv{\Phi} \hmv{x}(t, \mv{\mu}),
\label{eq:low-dimensional continuous state representation} \\
\dot{\tmv{x}}(t, \mv{\mu}) &= \mv{\Phi} \dot{\hmv{x}}(t, \mv{\mu}), \label{eq:low-dimensional velocity representation} 
\end{align}
respectively, with ${\tilde{\mv{x}}}, \dot{\tilde{\mv{x}}}: [0,T] \times
\RR^{n_{\mu}} \to \RR^N$; $\xref:\mv{\mu}\rightarrow \RR^N$ denoting a
reference state; and $\hmv{x}: [0,T] \times \RR^{n_{\mu}} \to \RR^p$ and
$\dot{\hmv{x}}: [0,T] \times \RR^{n_{\mu}} \to \RR^p$ denoting the
time-continuous generalized coordinates for the state and velocity,
respectively.
The time-discrete ROM state then satisfies
\begin{align}
\tmv{x}^{n}(\mv{\mu}) &\coloneqq \mv{x}_{\mathrm{ref}}(\mv{\mu}) + \mv{\Phi}
	\hmv{x}^{n}(\mv{\mu}),\quad n=1,\ldots,N_T \label{eq:low dimensional discrete state representation}
\end{align}
with 
$\tmv{x}^{n}:\RR^{n_{\mu}}\to \RR^N$, $n=1,\ldots,N_T$
and $\hmv{x}^{n}: 
\RR^{n_{\mu}} \to \RR^p$, $n=1,\ldots,N_T$ denoting the time-discrete
generalized coordinates for the state.
One
convenient choice for the reference state comprises $\xref(\mv{\mu}) =
\mv{x}^0(\mv{\mu})$, which implies that the initial condition can be exactly
satisfied by setting $\hmv{x}(0, \mv{\mu}) = \hmv{x}^{0}(\mv{\mu}) = \mv{0}$.

\subsection{Galerkin projection}
\label{sec:Galerkin projection reduced-order model}

Given the kinematic approximation \eqref{eq:low-dimensional continuous state
representation}--\eqref{eq:low dimensional discrete state representation},
Galerkin projection derives the dynamics of the ROM by setting the
(time-continuous) velocity to be the minimizer of the time-continuous residual \eqref{eq:time-continuous full-order residual} over the trial
subspace given the current state, time, and parameters, i.e., 
\begin{equation}
\label{eq:unconstrained Galerkin projection optimization problem}
\dot{\hmv{x}} \in \arg \min_{\hmv{v} \in \mathbb{R}^p} \norm{\mv{r}( \mv{\Phi} \hmv{v}, \xref(\mv{\mu}) + \mv{\Phi} \hmv{x}, t; \mv{\mu})}_2.
\end{equation}
One can show that necessary stationarity conditions for the solution to
\eqref{eq:unconstrained Galerkin projection optimization problem} are
equivalent to the familiar low-dimensional Galerkin ODE system
\begin{equation}\label{eq:galerkin ODE}
\dot{\hmv{x}} = \mv{\Phi}^T \mv{f}(\xref(\mv{\mu}) + \mv{\Phi} \hmv{x}, t; \mv{\mu}).
\end{equation}
Application of a linear multistep scheme to numerically solve equation
\eqref{eq:galerkin ODE} yields the time-discrete Galerkin O$\Delta$E
\begin{equation}
\label{eq:unconstrained Galerkin projection residual}
\hmv{r}^{n}_G(\xhat^{n}; \mv{\mu}) = \mv{0},\quad
	n=1,\ldots,N_T,
\end{equation}
where
\begin{equation}
\begin{split}
\hmv{r}^{n}_G&: (\hmv{\xi}; \mv{\nu}) \mapsto \alpha_0 \hmv{\xi} - \Delta t \beta_0 \mv{\Phi}^T \mv{f}(\xref(\mv{\nu}) + \mv{\Phi} \hmv{\xi}, t^{n}; \mv{\nu}) \\ 
&+ \sum_{j=1}^k \alpha_i \xhat^{n-j} - \Delta t \sum_{j=1}^k \beta_j
	\mv{\Phi}^T \mv{f}(\xref(\mv{\nu}) + \mv{\Phi} \xhat^{n-j}, t^{n-j};
	\mv{\nu}),\quad n=1,\ldots,N_T
\end{split}
\end{equation}
with $\hmv{r}^{n}_G: \RR^p \times \RR^{n_{\mu}} \to \RR^p$,
$n=1,\ldots,N_T$ denoting the time-discrete Galerkin residual.
Solving equation \eqref{eq:unconstrained Galerkin projection residual} yields
the parameterized time-discrete Galerkin state
$\hmv{x}^{n}(\mv{\mu})$, $n=1,\ldots,N_T$.

\subsection{Least-squares Petrov--Galerkin projection}
\label{sec:Least-squares Petrov--Galerkin projection reduced-order model}
Given the kinematic approximation \eqref{eq:low-dimensional continuous state
representation}--\eqref{eq:low dimensional discrete state representation}, LSPG projection \cite{carlberg2017galerkin} derives the ROM dynamics by
posing a time-discrete optimization problem. That is, it computes the
time-discrete ROM state as the minimizer of the 
time-discrete residual \eqref{eq:linear multistep scheme} such that
\begin{equation}
\label{eq:unconstrained LSPG optimization problem}
\begin{split}
\hmv{x}^{n} \in  &\arg \min_{\hmv{\xi} \in \mathbb{R}^p}
	\norm{\mv{r}^{n}(\xref(\mv{\mu}) + \mv{\Phi} \hmv{\xi}; \mv{\mu})}_2,\quad
	n=1,\ldots,N_T.
\end{split}
\end{equation}
One can show that necessary stationarity conditions for the solution to  
\eqref{eq:unconstrained LSPG optimization problem} correspond to the LSPG
O$\Delta$E
\begin{equation}
\hmv{r}^{n}_L(\hmv{x}^{n}; \mv{\mu}) = \mv{0},\quad n=1,\ldots,N_T,
\end{equation} 
where
\begin{equation}
\hmv{r}_L^{n}: (\hmv{\xi}; \mv{\nu}) \mapsto
\mv{\Phi}^T 
	\left[\frac{\partial \mv{r}^{n}}{\partial \mv{\xi}}(\xref(\mv{\nu}) + \mv{\Phi} \hmv{\xi}; \mv{\nu})\right]^T
	\mv{r}^{n}(\xref(\mv{\nu}) + \mv{\Phi} \hmv{\xi}; \mv{\nu}),\quad
	n=1,\ldots,N_T
\end{equation}
with $\hmv{r}^{n}_L: \RR^p \times \RR^{n_{\mu}} \to \RR^p$, $n=1,\ldots,N_T$
denoting the time-discrete LSPG O$\Delta$E residual. 

\section{Constrained Galerkin and LSPG projection}
\label{sec:Constrained optimization-based dimensional reduction}
This section introduces the proposed constrained Galerkin and LSPG projection
techniques.
We begin by defining general kinematic and dynamic equality and inequality
constraints
in Section \ref{sec:General constraint formulation}. Section
\ref{sec:Constrained Galerkin projection reduced-order model} and
\ref{sec:Constrained least-squares Petrov--Galerkin reduced-order model}
derive the Galerkin and LSPG projection techniques, respectively.

\subsection{General constraint formulation}
\label{sec:General constraint formulation}
 We introduce a general set of constraints 
\begin{align}
\bmv{c}(\mv{x}, t; \mv{\mu}) &= \mv{0}, \quad \forall t \in [0, T],\ \forall
	\mv{\mu}\in\mathcal D, \label{eq:kinematic equality constraint} \\
\mv{c}(\dmv{x}, \mv{x}, t; \mv{\mu}) &= \mv{0}, \quad \forall t \in [0, T],\ \forall
	\mv{\mu}\in\mathcal D, \label{eq:dynamic equality constraint} \\
\bmv{d}(\mv{x}, t; \mv{\mu}) &\geq \mv{0}, \quad \forall t \in [0, T],\ \forall
	\mv{\mu}\in\mathcal D, \label{eq:kinematic inequality constraint} \\
\mv{d}(\dmv{x}, \mv{x}, t; \mv{\mu}) &\geq \mv{0}, \quad \forall t \in [0, T],\ \forall
	\mv{\mu}\in\mathcal D, \label{eq:dynamic inequality constraint}
\end{align}
where $\geq \mv{0}$ denotes componentwise inequality. Equations
\eqref{eq:kinematic equality constraint}-\eqref{eq:dynamic inequality
constraint} can be categorized as either equality or inequality constraints and
as either kinematic constraints on the state or dynamic constraints on the
velocity (given the state), i.e.,
\begin{itemize}
\item kinematic equality constraints 
\begin{align}
\begin{split}
\bmv{c}: 
&(\mv{\xi}, \tau; \mv{\nu}) \mapsto \bmv{c}(\mv{\xi}, \tau; \mv{\nu}) \\
:&\RR^N \times [0,T] \times \RR^{n_{\mu}} \to \RR^{n_{\bar{c}}},
\end{split}
\end{align}
\item dynamic equality constraints 
\begin{align}
\begin{split}
\mv{c}: 
&(\mv{v}, \mv{\xi}, \tau; \mv{\nu}) \mapsto \mv{c}(\mv{v}, \mv{\xi}, \tau; \mv{\nu}) \\
:&\RR^N \times \RR^N \times [0,T] \times \RR^{n_{\mu}} \to \RR^{n_{c}},
\end{split}
\end{align}
\item kinematic inequality constraints 
\begin{align}
\begin{split}
\bmv{d}: 
&(\mv{\xi}, \tau; \mv{\nu}) \mapsto \bmv{d}(\mv{\xi}, \tau; \mv{\nu}) \\
:&\RR^N \times [0,T] \times \RR^{n_{\mu}} \to \RR^{n_{\bar{d}}},
\end{split}
\end{align}
\item dynamic inequality constraints
\begin{align}
\begin{split}
\mv{d}: 
&(\mv{v}, \mv{\xi}, \tau; \mv{\nu}) \mapsto \mv{d}(\mv{v}, \mv{\xi}, \tau; \mv{\nu}) \\
:&\RR^N \times \RR^N \times [0,T] \times \RR^{n_{\mu}} \to \RR^{n_{d}}.
\end{split}
\end{align}
\end{itemize}
Specific instances representable by our general constraint set can be found in literature. Table \ref{tab:literature overview} states examples of such constraints.

\begin{remark}
The purpose of this work is to augment standard minimum-residual reduced-order models with general kinematic and dynamic constraints to enforce structure preservation or other desired properties. However, it is possible to add constraints that render the optimization problem infeasible, which can happen if the optimization problem becomes overconstrained or the reduced basis
precludes feasibility, i.e., there is no solution in the reduced subspace that satisfies the constraints. Both of these issues are problem-dependent and are beyond the scope of this work.
\end{remark}

\subsection{Constrained Galerkin projection}
\label{sec:Constrained Galerkin projection reduced-order model}
As can be seen from equation \eqref{eq:unconstrained Galerkin projection
optimization problem}, Galerkin projection corresponds to solving a
time-continuous
optimization problem for the velocity generalized coordinates. As such, we
must linearize the kinematic constraints such that they are enforced to first
order by the computed velocity.
The associated linearizations of constraints
\eqref{eq:kinematic equality constraint} and \eqref{eq:kinematic inequality
constraint} are
\begin{align}
\frac{\partial \bmv{c}}{\partial \mv{\xi}}(\mv{x}, t; \mv{\mu}) \dmv{\mv{x}} +
	\frac{\partial \bmv{c}}{\partial \tau}(\mv{x}, t; \mv{\mu}) &= \mv{0},
	\hspace{94pt} \forall t \in [0, T],\ \forall \mv{\mu}\in\mathcal D, \label{eq:kinematic equality constraint total derivative} \\
\frac{\partial \bmv{d}}{\partial \mv{\xi}}(\mv{x}, t; \mv{\mu}) \dmv{\mv{x}} + \frac{\partial \bmv{d}}{\partial \tau}(\mv{x}, t; \mv{\mu}) &\geq \mv{0} \qquad \mathrm{if} \ \bmv{d}(\mv{x}, t; \mv{\mu}) = \mv{0}, \quad \forall t \in [0, T],\ \forall \mv{\mu}\in\mathcal D, \label{eq:kinematic inequality constraint total derivative}
\end{align}
respectively. Note that the condition on inequality \eqref{eq:kinematic
inequality constraint total derivative} arises from the need to enforce the
constraint to first order only if the constraint is active.

Recalling the definitions of the ROM state
\eqref{eq:low-dimensional continuous state representation} and velocity \eqref{eq:low-dimensional velocity representation}, we introduce the set of general Galerkin projection constraints
\begin{align}
\bmv{c}_G&: (\hmv{v}, \hmv{\xi}, \tau; \mv{\nu}) \mapsto \frac{\partial \bmv{c}}{\partial \mv{\xi}}(\xref(\mv{\nu}) + \mv{\Phi} \hmv{\xi}, \tau; \mv{\nu}) \mv{\Phi} \hmv{v} + \frac{\partial \bmv{c}}{\partial \tau}(\xref(\mv{\nu}) + \mv{\Phi} \hmv{\xi}, \tau; \mv{\nu}), \\
\mv{c}_G&: (\hmv{v}, \hmv{\xi}, \tau; \mv{\nu}) \mapsto \mv{c}(\mv{\Phi} \hmv{v}, \xref(\mv{\nu}) + \mv{\Phi} \hmv{\xi}, \tau; \mv{\nu}), \\
\bmv{d}_G&: (\hmv{v}, \hmv{\xi}, \tau; \mv{\nu}) \mapsto \frac{\partial \bmv{d}}{\partial \mv{\xi}}(\xref(\mv{\nu}) + \mv{\Phi} \hmv{\xi}, \tau; \mv{\nu}) \mv{\Phi} \hmv{v} + \frac{\partial \bmv{d}}{\partial \tau} (\xref(\mv{\nu}) + \mv{\Phi} \hmv{\xi}, \tau; \mv{\nu}), 
\label{eq:Galerkin kinematic inequality constraint} \\
\mv{d}_G&: (\hmv{v}, \hmv{\xi}, \tau; \mv{\nu}) \mapsto  \mv{d}(\mv{\Phi} \hmv{v}, \xref(\mv{\nu}) + \mv{\Phi} \hmv{\xi}, \tau; \mv{\nu}), \label{eq:Galerkin dynamic inequality constraint}
\end{align}
with $\bmv{c}_G, \mv{c}_G, \bmv{d}_G, \mv{d}_G: \RR^p \times \RR^p \times [0, T]
\times \RR^{n_{\mu}} \to \RR^{\gamma}$, $\gamma \in \{n_{\bar{c}}, n_{c},
n_{\bar{d}}, n_d\}$ all velocity dependent.

As a result, the time-continuous constrained Galerkin ODE system is
\begin{equation}
\dot{\xhat} = \hmv{f}(\hmv{x}, t; \mv{\mu})
\label{eq:constrained Galerkin projection ROM}
\end{equation}
with the velocity $\hmv{f}: (\hmv{\xi}, \tau; \mv{\nu}) \mapsto \hmv{f}(\hmv{\xi}, \tau;
\mv{\nu})$, $\hmv{f}: \mathbb{R}^p \times [0, T] \times \mathbb{R}^{n_{\mu}}
\to \mathbb{R}^p$ corresponding to the solution of the optimization problem
\begin{equation}
\begin{split}
\label{eq:constrained Galerkin projection optimization problem}
&\minimize{\hmv{v} \in \mathbb{R}^p} \norm{\mv{r}( \mv{\Phi} \hmv{v}, \xref(\mv{\nu}) + \mv{\Phi} \hmv{\xi}, \tau; \mv{\nu})}_2 \\
&\mathrm{subject \ to} \qquad \bmv{c}_G(\hmv{v}, \hmv{\xi}, \tau; \mv{\nu}) = \mv{0} \\
&\hspace{60pt} \mv{c}_G(\hmv{v}, \hmv{\xi}, \tau; \mv{\nu}) = \mv{0} \\
&\hspace{59pt} \bmv{d}_G(\hmv{v}, \hmv{\xi}, \tau; \mv{\nu}) \geq \mv{0} \qquad \mathrm{if} \ \bmv{d}(\xref(\mv{\nu}) + \mv{\Phi} \hmv{\xi}, \tau; \mv{\nu}) = \mv{0} \\
&\hspace{59pt} \mv{d}_G(\hmv{v}, \hmv{\xi}, \tau; \mv{\nu}) \geq \mv{0}.
\end{split}
\end{equation}

The resulting time-discretized constrained Galerkin O$\Delta$E is
\begin{equation}
\label{eq:time-discretized Galerkin projection in residual notation}
\mv{r}^{n}_G(\xhat^{n}; \mv{\mu}) = \mv{0},\quad n=1,\ldots,N_T,
\end{equation}
where 
\begin{equation}
\label{eq:constrained Gelerkin projection residual}
\mv{r}^{n}_G: (\hmv{\xi}; \mv{\nu}) \mapsto \alpha_0 \hmv{\xi} - \Delta t \beta_0
	\hmv{f}(\hmv{\xi}, t^{n}; \mv{\nu}) + \sum_{j=1}^k \alpha_j \xhat^{n-j} -
	\Delta t \sum_{j=1}^k \beta_j \hmv{f}(\xhat^{n-j}, t^{n-j}; \mv{\nu}),\quad
	n=1,\ldots,N_T
\end{equation}
with $\mv{r}^{n}_G: \RR^{p} \times \RR^{n_{\mu}} \to \RR^{p}$,
$n=1,\ldots,N_T$ denoting the constrained Galerkin O$\Delta$E
residual.

\begin{remark}
\label{remark:kinematic inequality constraint for Galerkin projection}
The conditional statement ``$\mathrm{if} \ \bmv{d}(\xref(\mv{\nu}) + \mv{\Phi}
	\hmv{\xi}, \tau; \mv{\nu}) = \mv{0}$'' in optimization problem
	\eqref{eq:constrained Galerkin projection optimization problem} 
	is employed to identify active constraints from the time-continuous
	perspective. In practice, strict fulfillment of
	the associated kinematic inequality constraint can only be
	guaranteed for infinitesimally small time steps. Once time discretization is
	introduced in
	\eqref{eq:time-discretized Galerkin projection in residual
	notation}, finite time steps and limited machine precision imply that strict
	satisfaction of the kinematic inequality will likely fail in practice. To
	address this, we trigger
	activation of the constraint by admitting potential violation; that is, the
	conditional statement in optimization problem \eqref{eq:constrained Galerkin
	projection optimization problem} after time discretization reads
	``$\mathrm{if} \ \bmv{d}(\xref(\mv{\nu}) + \mv{\Phi} \hmv{\xi}, \tau;
	\mv{\nu}) \leq \mv{0}$''. As a consequence, slight constraint violations due
	to time discretization and finite precision will be accepted by the
	numerical solution.
\end{remark}

\subsection{Constrained least-squares Petrov--Galerkin projection}
\label{sec:Constrained least-squares Petrov--Galerkin reduced-order model}
The LSPG projection ROM \eqref{eq:unconstrained LSPG optimization problem}
corresponds to solving a time-discrete optimization problem for the state
generalized coordinates. As such, we must discretize the dynamic constraints
to convert them into constraints that depend on the state only. To achieve
this, we apply the same underlying time-discretization scheme used to
discretize the velocity in the governing ODE. In the
case of linear multistep schemes, the discretized dynamic constraints
can be derived by substituting the time-continuous velocity with its
time-discrete variant 
\begin{equation}
\bmv{v}^{n-q}:
(\mv{\xi}; \mv{\nu}) \mapsto \frac{1}{\Delta t \beta_q} ( \alpha_0 \mv{\xi} +
	\sum_{j=1}^{k} \alpha_j \mv{x}^{n-j} - \Delta t \sum_{j=1+q}^{k} \beta_j
	\mv{f}(\mv{x}^{n-j}, t^{n-j}; \mv{\nu}) ),\quad n=1,\ldots,N_T
\end{equation}
with $\bmv{v}^{n-q}: \RR^N \times \RR^{n_{\mu}} \to \RR^N$ and $q=0$ for $\beta_0 \neq 0$ (implicit time integration) and $q=1$ for $\beta_0 = 0$ (explicit time integration). As a consequence, we can express the set of general LSPG constraints as  
\begin{align}
\bmv{c}_{L}^{n}&: (\hmv{\xi}; \mv{\nu}) \mapsto \bmv{c}(\xref(\mv{\nu}) + \mv{\Phi} \hmv{\xi}, t^{n}; \mv{\nu}), \\ 
\mv{c}_{L}^{n}&: (\hmv{\xi}; \mv{\nu}) \mapsto \mv{c}(\bmv{v}^{n-q}(\xref(\mv{\nu}) + \mv{\Phi} \hmv{\xi}; \mv{\nu}), \xref(\mv{\nu}) + \mv{\Phi} \hmv{\xi}^n_q, t^{n-q}; \mv{\nu}), \label{eq:LSPG dynamic equality constraint} \\
\bmv{d}_{L}^{n}&: (\hmv{\xi}; \mv{\nu}) \mapsto \bmv{d}(\xref(\mv{\nu}) + \mv{\Phi} \hmv{\xi}, t^{n}; \mv{\nu}), \label{eq:LSPG kinematic inequality} \\ 
\mv{d}_{L}^{n}&: (\hmv{\xi}; \mv{\nu}) \mapsto \mv{d}(\bmv{v}^{n-q}(\xref(\mv{\nu}) + \mv{\Phi} \hmv{\xi}; \mv{\nu}), \xref(\mv{\nu}) + \mv{\Phi} \hmv{\xi}^n_q, t^{n-q}; \mv{\nu}) \label{eq:LSPG dynamic inequality constraint},
\end{align}
with $\bmv{c}_{L}^{n}, \mv{c}_{L}^{n}, \bmv{d}_{L}^{n}, \mv{d}_{L}^{n}: \RR^p \times \RR^{n_{\mu}} \to \RR^{\gamma}, \ \gamma \in \{n_{\bar{c}}, n_c, n_{\bar{d}}, n_d\}$ and $q=0$ for $\beta_0 \neq 0$ and $q=1$ for $\beta_0 = 0$. For notational convenience we introduced
\begin{equation}
\hmv{\xi}^n_q =
\left \{
\begin{aligned}
\hmv{\xi} \quad &\mathrm{for} \ q = 0 \\
\hmv{x}^{n-1} \quad &\mathrm{for} \ q = 1,
\end{aligned}
\right.
\end{equation}
resolving either to the optimization variable $\hmv{\xi}$ or to the fixed generalized-coordinates state of the previous time step $\hmv{x}^{n-1}$.

Then, the time-discrete LSPG ROM solution $\xhat^{n}$,
$n=1,\ldots,N_T$ corresponds to the solution to the optimization
problem 
\begin{equation}
\label{eq:LSPG optimization problem}
\begin{split}
&\minimize{\hmv{\xi} \in \mathbb{R}^p} \norm{\mv{r}^{n}(\xref(\mv{\mu}) + \mv{\Phi} \hmv{\xi}; \mv{\mu})}_2 \\
&\mathrm{subject \ to} \quad \bmv{c}_{L}^{n}(\hmv{\xi}; \mv{\mu}) = \mv{0} \\
&\hspace{50pt} \mv{c}_{L}^{n}(\hmv{\xi}; \mv{\mu}) = \mv{0} \\
&\hspace{49pt} \bmv{d}_{L}^{n}(\hmv{\xi}; \mv{\mu}) \geq \mv{0} \\
&\hspace{49pt} \mv{d}_{L}^{n}(\hmv{\xi}; \mv{\mu}) \geq \mv{0}. \\
\end{split}
\end{equation}

\begin{remark}
The introduction of the state-to-velocity mapping into the dynamic constraints \eqref{eq:LSPG dynamic equality constraint} and \eqref{eq:LSPG dynamic inequality constraint} is consistent with the time discretization of the FOM. As such, the time-discrete residual \eqref{eq:linear multistep scheme} can be written in terms of the time-continuous residual \eqref{eq:time-continuous full-order residual} as
\begin{equation}
\mv{r}^{n}(\mv{x}^{n}; \mv{\mu}) = 
\Delta t \beta_q \mv{r}(\bmv{v}^{n-q} (\mv{x}^{n}; \mv{\mu}), \mv{x}^{n-q}, t^{n-q}; \mv{\mu}),
\end{equation}
with $q=0$ for $\beta_0 \neq 0$ and $q=1$ for $\beta_0 = 0$, compare with time discretization of the dynamic constraints \eqref{eq:LSPG dynamic equality constraint} and \eqref{eq:LSPG dynamic inequality constraint}.
\end{remark} 

\begin{remark}
The unconstrained projections require solving a nonlinear system of size $p$,
	where the dominant cost comes from evaluating the reduced residual and
	Jacobian and solving the (dense) linear system at each nonlinear iteration.
	In contrast, the constrained projections require solving an optimization
	problem in $p$ variables and $M=n_{\bar{c}}+n_c+n_{\bar{d}}+n_d$
	constraints. The dominant cost comes from evaluating the residual and its
	Jacobian, evaluating the constraints and their Jacobian; the system that
	must be solved (which is usually dense) depends on the optimization solver
	used. For example, for penalty and augmented Lagrangian methods, the
	constrained optimization problem is solved by solving a sequence of
	unconstrained optimization problems whose objective incorporates the
	original objective function and constraints. The unconstrained problem is
	usually solved using quasi-Newton methods, which results in a dense linear
	system of size $p$. The cost per iteration of such methods for the
	constrained projection is similar to a nonlinear iteration for the
	unconstrained projection (depending on the cost of constraint evaluations);
	however, the constrained problem usually requires more iterations to
	converge. On the other hand, sequential quadratic programming and
	primal-dual interior point methods result in higher-dimensional linear
	systems, which increases the cost per iteration, but they usually converge more rapidly. In this work, $M\leq 3$ so the increased cost per iteration (relative to the unconstrained projection) is negligible.
\end{remark}

\begin{remark}
The main step required to incorporate hyper-reduction into the constrained model reduction formulation is acceleration of the discrete residual and Jacobian evaluations, which can be achieved using standard techniques \cite{an2008optimizing,chaturantabut2010nonlinear,Farhat2014625}. This is the only step required if the constraints can be evaluated at a cost independent of the FOM dimension $N$ (or reduces as such when the model order reduction ansatz is used, e.g., polynomial in $\mv{x}$ and $\dot{\mv{x}}$). In the case where the constraint evaluations scale with $N$ (e.g., non-polynomial dependence on $\mv{x}$ and $\dot{\mv{x}}$), hyper-reduction must be incorporated into the constraint functions; see Ref.~\cite{balajewicz2016projection} in the context of contact mechanics and
Ref.~\cite{Carlberg2018280} in the context of fluid mechanics. A general procedure to hyper-reduce constraint functions will be considered in future work.
\end{remark}

\section{Specific constraints}
\label{sec:Specific constraints}
The general formulation of kinematic/dynamic equality/inequality constraints
presented in Section \ref{sec:General constraint formulation} can be employed
as a template for defining problem-specific constraints. In this section we
derive such specific constraints, which will be enforced in the numerical
examples in Section \ref{sec:Numerical experiments}.

\subsection{Constraint on the sum of residual entries (dynamic equality)}
By definition, the time-continuous FOM solution satisfies constraints
\begin{equation}
\mv{C}\mv{r}(\dot{\mv{x}}, \mv{x}, t; \mv{\mu}) = \mv{0}
\label{eq:time-continuous rsum-constraint}
\end{equation}
and the time-discrete FOM solution satisfies
\begin{equation}
\mv{C}\mv{r}^{n}(\mv{x}^{n}; \mv{\mu}) = 
\Delta t \beta_q \mv{C}\mv{r}(\bmv{v}^{n-q} (\mv{x}^{n}; \mv{\mu}), \mv{x}^{n-q}, t^{n-q}; \mv{\mu})
=
	\mv{0},\quad n=1,\ldots,N_T
\label{eq:time-discrete rsum-constraint}
\end{equation}
for any matrix $\mv{C} \in \RR^{n_{C} \times N}$. This type of dynamic equality constraint can be
employed to enforce conservation in the case of FOMs associating with
finite-volume discretizations of conservation laws \cite{Carlberg2018280}.
We also consider this type of constraint matrix, and particularly focus on
cases where the matrix is composed exclusively of 0 and 1 entries. In this
case, equation
\eqref{eq:time-continuous rsum-constraint} and \eqref{eq:time-discrete
rsum-constraint} return a vector of sums of selected residual entries, in the
following we will refer to this type of constraint as a
sum-of-residual-entries (rsum) constraint.

The Galerkin projection constraint corresponding to optimization problem \eqref{eq:constrained Galerkin projection optimization problem} reads
\begin{equation}
\label{eq:Galerkin conservation constraint}
\mv{c}_{G, \mathrm{rsum}}(\hmv{v}, \hmv{\xi}, \tau; \mv{\nu}) = \mv{0}
\end{equation}
using
\begin{equation}
\mv{c}_{G, \mathrm{rsum}}: (\hmv{v}, \hmv{\xi}, \tau; \mv{\nu}) \mapsto \mv{C} \mv{r}( \mv{\Phi} \hmv{v}, \xref(\mv{\nu}) + \mv{\Phi} \hmv{\xi}, \tau; \mv{\nu})
\end{equation}
with $\mv{c}_{G, \mathrm{rsum}}: \RR^p \times \RR^p \times [0, T] \times \RR^{n_{\mu}} \to \RR^{n_{C}}$.

The LSPG projection constraint corresponding to optimization problem \eqref{eq:LSPG optimization problem} reads
\begin{equation}
\label{eq:LSPG conservation constraint}
\mv{c}^{n}_{L, \mathrm{rsum}}(\hmv{\xi}; \mv{\mu}) = \mv{0},\quad n=1,\ldots,N_T
\end{equation}
using 
\begin{equation}
\mv{c}^{n}_{L, \mathrm{rsum}}(\hmv{\xi}; \mv{\nu}) \mapsto \mv{C}
	\mv{r}^{n}(\xref(\mv{\nu}) + \mv{\Phi} \hmv{\xi}; \mv{\nu}) = 
\Delta t \beta_q \mv{C}\mv{r}(\bmv{v}^{n-q} (\xref(\mv{\nu}) + \mv{\Phi} \hmv{\xi}; \mv{\nu}), \mv{x}^{n-q}, t^{n-q}; \mv{\nu})
\end{equation}
with $\mv{c}^{n}_{L, \mathrm{rsum}}: \RR^p \times \RR^{n_{\mu}} \to
\RR^{n_{C}}$, $n=1,\ldots,N_T$. 

\subsection{Constraint on diminishing total variation (dynamic inequality)}
Another useful property that holds for scalar conservation laws is
diminishing total variation (referred to as tvd-constraint in the following).
For the case of a 1D problem, the total variation of the solution---in the
case of the state $\mv{x}$ corresponding to the value of the solution on a 1D
uniformly spaced grid---can be expressed as 
\begin{equation}
\label{eq:total variation definition}
TV(\mv{x}) = \sum^{N-1}_{i=1} | [\mv{x}]_{i+1} - [\mv{x}]_i |,
\end{equation}
where $[\mv{x}]_i$ denotes the $i$th component of the vector $\mv{x}$.
Non-increasing total variation can be expressed as the dynamic inequality
constraint
\begin{equation}
\frac{d }{d t} TV(\mv{x}) \leq 0,
\end{equation}
which is equivalent to
\begin{equation}
- \sum^{N-1}_{i=1} \sgn([\mv{x}]_{i+1} - [\mv{x}]_i)([\dot{\mv{x}}]_{i+1} - [\dot{\mv{x}}]_{i}) \geq 0,
\label{eq:definition tvd-constraint}
\end{equation}
wherein we introduced the sign-function $\sgn(x) = \frac{d}{dx} |x| \  \mathrm{for} \ x \neq 0$.
The Galerkin projection constraint corresponding to optimization problem \eqref{eq:constrained Galerkin projection optimization problem} reads
\begin{equation}
d_{G, \mathrm{tvd}}(\hmv{v}, \hmv{\xi}, \bullet; \mv{\nu}) \geq 0
\label{eq:Galerkin TVD constraint}
\end{equation}
using
\begin{equation}
\begin{split}
d_{G, \mathrm{tvd}}: (\hmv{v}, \hmv{\xi}, \bullet; \mv{\nu}) \mapsto - \sum^{N-1}_{i=1} \sgn([\xref(\mv{\nu}) + \mv{\Phi} \hmv{\xi}]_{i+1} - [\xref(\mv{\nu}) + \mv{\Phi} \hmv{\xi}]_i)
([\mv{\Phi} \hmv{v}]_{i+1} - [\mv{\Phi} \hmv{v}]_{i})
\end{split}
\end{equation}
with $d_{G, \mathrm{tvd}}: \RR^{p} \times \RR^{p} \times \RR^{n_{\mu}} \to \RR$ and $\bullet$ indicating an unused argument of the template constraint \eqref{eq:dynamic inequality constraint}.

The LSPG projection constraint corresponding to optimization problem \eqref{eq:LSPG optimization problem} reads
\begin{equation}
\label{eq:LSPG TVD constraint}
d^{n}_{L, \mathrm{tvd}}(\hmv{\xi}; \mv{\nu})  \geq 0,\quad
	n=1,\ldots,N_T
\end{equation}
using
\begin{equation}
\begin{split}
d^{n}_{L, \mathrm{tvd}}(\hmv{\xi}; \mv{\nu}) \mapsto - \sum^{N-1}_{i=1} &\sgn([\xref(\mv{\nu}) + \hmv{\xi}^{n}_q]_{i+1} - [\xref(\mv{\nu}) + \hmv{\xi}^{n}_q]_i) \\ 
&([\bmv{v}^{n-q}(\xref(\mv{\nu}) + \mv{\Phi} \hmv{\xi}; \mv{\nu})]_{i+1} - [\bmv{v}^{n-q}(\xref(\mv{\nu}) + \mv{\Phi} \hmv{\xi}; \mv{\nu})]_{i})
\end{split} 
\end{equation}
with $d^{n}_{L, \mathrm{tvd}}: \RR^p \times \RR^{n_{\mu}} \to \RR$,
$n=1,\ldots,N_T$. 

\subsection{Constraint on total variation bound (kinematic inequality)}
In some cases, schemes enforce a bound on the total variation rather than
enforcing a strict total-variation-diminishing constraint due to local
degeneracy to first order at critical points. Assuming a given
bound $b$ is enforced on the total variation, a total variation bounding constraint (referred to as tvb-constraint in the following)---in the
case of the state $\mv{x}$ corresponding to the value of the solution on a 1D
uniformly spaced grid---reads
\begin{equation}
TV(\mv{x}) \leq b.
\label{eq:tvb-constraint}
\end{equation} 

As a consequence, the Galerkin projection constraint corresponding to optimization problem \eqref{eq:constrained Galerkin projection optimization problem} reads
\begin{equation}
\label{eq:Galerkin tvb-constraint}
\bar{d}_{G, \mathrm{tvd}}(\hmv{v}, \hmv{\xi}, \bullet; \mv{\nu}) \geq 0 \quad \mathrm{if} \ TV(\xref(\mv{\nu}) + \mv{\Phi} \hmv{\xi}) - b = 0.
\end{equation}

The LSPG projection constraint corresponding to optimization problem \eqref{eq:LSPG optimization problem} reads
\begin{equation}
\label{eq:LSPG tvb-constraint}
\bar{d}_{L, \mathrm{tvb}}^{n}(\hmv{\xi}; \mv{\nu}) \geq 0,\quad
	n=1,\ldots,N_T
\end{equation}
using
\begin{equation}
\bar{d}_{L, \mathrm{tvb}}^{n}: (\hmv{\xi}; \mv{\nu}) \mapsto b - TV(\xref(\mv{\nu}) + \mv{\Phi} \hmv{\xi}) 
\end{equation}
and $\bar{d}_{L, \mathrm{tvb}}^{n}: \RR^p \times \RR^{n_{\mu}} \to \RR$,
$n=1,\ldots,N_T$. 

\subsection{Constraint on conservation of energy (dynamic equality)}
We assume the existence of a history-independent  mapping $E$ returning the system energy $e$ from the current state
\begin{equation}
e = E(\mv{x})
\end{equation}
using
\begin{equation}
E: \mv{\xi} \mapsto E(\mv{\xi})
\end{equation}
 with $E: \RR^{N} \to \RR$.
Energy conservation (ec) is expressed as the dynamic equality constraint
\begin{equation}
\frac{\mathrm{d}}{\mathrm{d} t} E(\mv{x}) = S(t; \mv{\mu})
\end{equation}
or equivalently
\begin{equation}
\frac{\partial E}{\partial \mv{\xi}}(\mv{x})^T \dmv{x} - S(t; \mv{\mu}) = 0,
\label{eq:definition ec-constraint}
\end{equation}
wherein we introduced a state-independent source $S: [0, T] \times
\RR^{n_{\mu}} \to \RR$ (e.g., a heat source).

The Galerkin projection constraint corresponding to optimization problem \eqref{eq:constrained Galerkin projection optimization problem} reads
\begin{equation}
\label{eq:Galerkin projection ec-constraint}
c_{G,ec}(\hmv{v}, \hmv{\xi}, \tau; \mv{\nu}) = 0
\end{equation}
using 
\begin{equation}
c_{G,ec}: (\hmv{v}, \hmv{\xi}, \tau; \mv{\nu}) \mapsto \frac{\partial E}{\partial \mv{\xi}}(\xref(\mv{\nu}) + \mv{\Phi} \hmv{\xi})^T \mv{\Phi} \hmv{v} -  S(\tau; \mv{\nu})
\end{equation}
with $c_{G,ec}: \RR^p \times \RR^p \times [0,T] \times \RR^{n_{\mu}} \to \RR$.

The LSPG projection constraint corresponding to optimization problem \eqref{eq:LSPG optimization problem} reads
\begin{equation}
c_{L,ec}^{n}(\hmv{\xi}; \mv{\nu}) = 0,\quad n=1,\ldots,N_T
\label{eq:LSPG projection ec-constraint}
\end{equation}
using
\begin{equation}
c_{L,ec}^{n}: (\hmv{\xi}; \mv{\nu}) \mapsto \frac{\partial E}{\partial \mv{\xi}}(\xref(\mv{\nu}) + \mv{\Phi} \hmv{\xi}^n_q)^T \bmv{v}^{n-q}(\xref(\mv{\nu}) + \mv{\Phi} \hmv{\xi}; \mv{\nu}) - S(t^{n-q}; \mv{\nu})
\end{equation}
with $c_{L,ec}^{n}:\RR^p \times \RR^{n_{\mu}} \to \RR$ and $q=0$ for $\beta_0
\neq 0$ and $q=1$ for $\beta_0 = 0$ for $n=1,\ldots,N_T$.

\section{Numerical experiments}
\label{sec:Numerical experiments}
We now present several numerical examples demonstrating the performance and
flexibility of the
proposed formulation to enforce physical properties across a range of
problems. We begin by comparing a constrained and unconstrained
projection of FOM solutions onto a low-dimensional subspace. Next, we investigate the
performance of the proposed constrained Galerkin and LSPG projection methods
for a 1D Burgers' flow problem, a 1D Euler flow problem, and a 2D diffusion problem.

\subsection{Error metrics}
We employ several error metrics to assess the
discrepancy between the time-discrete FOM solution 
$\mv{x}^{n}$, $n=1,\ldots,N_T$
and the time-discrete ROM
solution
$\tmv{x}^{n}=\xref(\mv{\mu}) + \mv{\Phi} \hmv{x}^{n}$,
$n=1,\ldots,N_T$. In particular, we consider the following metrics:
\begin{itemize}

\item the relative state error at time step $n$: 
\begin{equation}
\varepsilon_x^{n} \coloneqq \norm{\mv{x}^{n} - (\xref(\mv{\mu}) + \mv{\Phi}
	\hmv{x}^{n})}_2 / \norm{\mv{x}^{n}}_2, \quad n=1,\ldots,N_T,
\label{eq:relative state space error}
\end{equation}

\item the violation of the Galerkin projection rsum-constraint at time step $n$:
\begin{equation}
\varepsilon_{G, \mathrm{rsum}}^{n} \coloneqq \norm{\mv{C} \mv{r}(\mv{\Phi} \hmv{f}(\xhat^{n}, t^{n}; \mv{\mu}), \xref(\mv{\mu}) + \mv{\Phi} \xhat^{n}, t^{n}; \mv{\mu})}_2, \quad n=1,\ldots,N_T,
\label{eq:violation of the Gelerkin projection rsum-constraint}
\end{equation}

\item the violation of the LSPG projection rsum-constraint at time step $n$:
\begin{equation}
\varepsilon_{L, \mathrm{rsum}}^{n} \coloneqq \norm{\mv{C} \mv{r}^{n}(\xref(\mv{\mu}) + \mv{\Phi}\hmv{x}^{n}; \mv{\mu})}_2, \quad n=1,\ldots,N_T,
\label{eq:violation of the LSPG projection rsum-constraint}
\end{equation}

\item the violation of the tvd-constraint at time step $n$:
\begin{equation}
\varepsilon_{\mathrm{tv}}^{n} \coloneqq \mathrm{max}(0, TV(\xref(\mv{\mu}) + \mv{\Phi}\hmv{x}^{n}) - TV(\xref(\mv{\mu}) + \mv{\Phi}\hmv{x}^{n-1})),\quad n=1,\ldots,N_T,
\label{eq:violation of tvd-constraint}
\end{equation}

\item the violation of the tvb-constraint at time step $n$:
\begin{equation}
\varepsilon_{\mathrm{tvb}}^{n} \coloneqq \max(0, TV(\xref(\mv{\mu}) + \mv{\Phi} \hmv{x}^{n}) - b),
\end{equation}
with the upper total variation bound $b$,

\item the absolute deviation of system energy from the initial condition $\mv{x}^0$ at time step $n$:
\begin{equation}
\varepsilon_{\mathrm{de}}^{n} \coloneqq |E(\xref(\mv{\mu}) + \mv{\Phi} \hmv{x}^{n}) - E(\mv{x}^0)|,\quad n=1,\ldots,N_T,
\label{eq:deviation of system energy}
\end{equation}

\item the global relative state error: 
\begin{equation}
\varepsilon_x \coloneqq \frac{\sqrt{\sum_{n=1}^{N_T} \norm{\mv{x}^n - (\xref(\mv{\mu}) + \mv{\Phi} \hmv{x}^{n})}^2_2}}{\sqrt{\sum_{n=1}^{N_T} \norm{\mv{x}^n}^2_2}}
\label{eq:integral relative state space error}
\end{equation}

\item the mean global violation of the Galerkin projection rsum-constraint:
\begin{equation}
\varepsilon_{G, \mathrm{rsum}} \coloneqq \frac{1}{N_T} \sqrt{\sum^{N_T}_{n=1} \left( \varepsilon^{n}_{G, \mathrm{rsum}} \right)^2},
\end{equation}

\item the mean global violation of the LSPG projection rsum-constraint:
\begin{equation}
\varepsilon_{L, \mathrm{rsum}} \coloneqq \frac{1}{N_T} \sqrt{\sum^{N_T}_{n=1} \left( \varepsilon^{n}_{L, \mathrm{rsum}} \right)^2},
\end{equation}

\item the mean global violation of the tvd-constraint:
\begin{equation}
\varepsilon_{\mathrm{tv}} \coloneqq \frac{1}{N_T} \sum^{N_T}_{n=1} \varepsilon_{\mathrm{tv}}^{n},
\end{equation}

\item the mean global violation of the tvb-constraint:
\begin{equation}
\varepsilon_{\mathrm{tvb}} \coloneqq \frac{1}{N_T} \sum^{N_T}_{n=1} \varepsilon_{\mathrm{tvb}}^n,
\end{equation}

\item and the mean global absolute deviation of system energy from the initial condition $\mv{x}^0$:
\begin{equation}
\varepsilon_{\mathrm{de}} \coloneqq \frac{1}{N_T} \sum^{N_T}_{n=1} \varepsilon_{\mathrm{de}}^{n}.
\end{equation}

\end{itemize}

\subsection{Nonlinear solution strategies}
\label{sec:Nonlinear solution strategies}

We apply the following solution methods for solving the nonlinear systems and
optimization problems that arise at each time step of the considered models.

\begin{itemize}
	\item FOMs are solved using Newton's method \newline
	(termination criterion: $\norm{\mv{r}^n(\mv{x}^n)}_2 < 10^{-10}$ or $\norm{\mv{r}^n(\mv{x}^n)}_2 / \norm{\mv{r}^n(\mv{x}^{n-1})}_2 < 10^{-6}$ ).

\item All constrained optimization problems are solved using gradient-based optimization methods with the objective gradient and constraint Jacobians derived analytically and implemented. In this work, we apply the general purpose optimization functionality provided by
		scipy.optimize.minimize (\textit{method}='SLSQP' and
		\textit{method}='trust-constr') available in the SciPy
		\cite{2020SciPy-NMeth} (version 1.3.0) scientific computing library of the
		Python programming language. For problem-specific optimizer settings see Section \ref{sec:Kinematics assessment: orthogonal projection vs. constrained projection} to \ref{sec:Diffusion equation}. 

\item The time-discretized Galerkin projection residual \eqref{eq:constrained
	Gelerkin projection residual} is defined implicitly, given that the velocity
		$\hmv{f}(\mv{\xi}, \tau; \mv{\nu})$ is the solution of optimization
		problem \eqref{eq:constrained Galerkin projection optimization problem}.
		In case of an implicit time stepping scheme, Newton's method would require
		the evaluation of residual derivatives, including differentiation of the
		velocity $\hmv{f}(\mv{\xi}, \tau; \mv{\nu})$. To avoid this, we apply
		derivative-free general purpose root finding functionality provided by
		scipy.optimize.root (\textit{method}='hybr' with parameters \textit{maxfev}=100, \textit{xtol}=$10^{-6}$) available in the SciPy
		scientific computing library mentioned above to solve the Galerkin
		O$\Delta$E \eqref{eq:time-discretized Galerkin projection in residual
		notation} in the case of implicit time integration. 
\end{itemize}

\subsection{Kinematics assessment: orthogonal projection vs. constrained projection}
\label{sec:Kinematics assessment: orthogonal projection vs. constrained projection}
In order to demonstrate the potential benefit of kinematic constraints alone, we
project the FOM solution onto a low-dimensional subspace 
using two projection approaches: (1) orthogonal $\ell^2$-projection, and (2)
kinematically constrained orthogonal projection using tvb-constraints. The arising optimization problems are solved by scipy.optimize.minimize functionality (\textit{method}='trust-constr' with parameters \textit{maxiter}=250, \textit{gtol}=$10^{-6}$, \textit{xtol}=$10^{-6}$, \textit{barrier\textunderscore tol}=$10^{-6}$, cf. Section \ref{sec:Nonlinear solution strategies}). 

\begin{figure}[h!]
\centering
\includegraphics[width=0.8\linewidth]{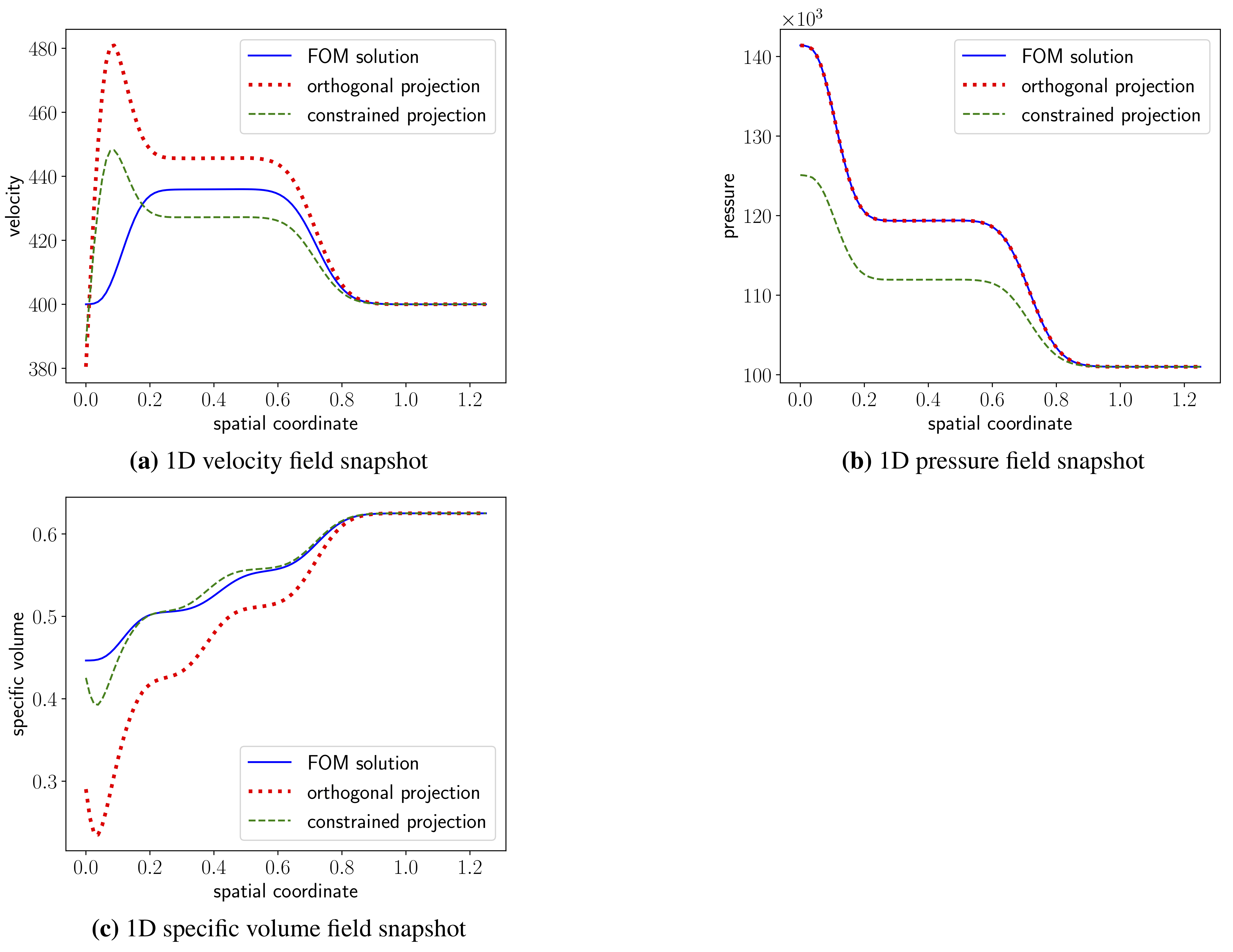} 
\caption{Exemplary FOM solution snapshot of 1D Euler flow, unconstrained orthogonal projection and tvb-constrained projection on the span of a given reduced-order basis.}
\label{fig:Euler snapshot projection}
\end{figure}

 Figure
\ref{fig:Euler snapshot projection} shows a solution snapshot $\mv{x}^{n} \in
\RR^{300} $ (blue curve) of the 1D Euler equations (at the parametric configuration $\mv{\mu} = (1.6, 1.4)$ and $n=200$, see Section \ref{sec:Euler equations} for the problem setup), incorporating the velocity field $u$, the pressure
field $p$ as well as the specific volume field $v$. Additionally, we plot the
orthogonal $\ell^2$-projection of this solution onto the affine subspace $\{\mv{\Phi}
\hmv{v} + \xref(\mv{\mu})\,|\,\hmv{v}\in\RR^{20}\}$
(red
curve) using a reduced-order basis $\mv{\Phi} \in 
V_{20}(\RR^{300})$ (which is computed from the first 20 POD modes of the FOM solution at parametric configuration $\mv{\mu} = (1.1, 1.1)$ ); the resulting solution
 $\hmv{x}^{n} \in \RR^{20}$ comprises the solution to the
optimization problem
\begin{equation} 
\label{eq:orthogonal projection Euler snapshot}
\minimize{\hmv{\xi} \in \RR^{20}} \norm{\mv{\Phi} \hmv{\xi} + \xref{\mv(\mv{\mu})} - \mv{x}^{n}}_2.
\end{equation}

\begin{remark}
The orthogonal-projection optimization problem \eqref{eq:orthogonal projection
	Euler snapshot} can be equivalently written as 
$\hmv{x}^{n} = \mv{\Phi}^T(\mv{x}^{n} - \xref(\mv{\mu}))$
due to orthogonality of 
the basis matrix $\mv{\Phi}$.
\end{remark}

Examination of Figure \ref{fig:Euler snapshot projection} reveals that the
absolute values of the pressure field are several orders of magnitude larger
than the velocity field, which in turn exhibits values several orders of
magnitude larger than the specific-volume field. As a consequence, the
orthogonal $\ell^2$-projection yields an approximation that accurately
reproduces the pressure field, but effectively neglects the velocity and specific-volume
field. 

We observe significant over- and undershoots of the
orthogonal projection for velocity and specific volume field. To avoid this,
we constrain the projection with a field-specific tvb-constraint of shape
\eqref{eq:LSPG tvb-constraint}, choosing the upper bound $b$ as 150\% of the
total variation of the FOM snapshot, that is, the tvb-constrained projection
reads
\begin{equation}
\begin{split}
&\minimize{\hmv{\xi} \in \mathbb{R}^{20}} \norm{\mv{\Phi} \hmv{\xi} + \xref(\mv{\mu}) - \mv{x}^{n}}_2 \\
&\mathrm{subject \ to} \quad \bar{c}_{L, \mathrm{tvb}}^{n, u}(\hmv{\xi}; \mv{\mu}) \geq 0 \\
&\hspace{50pt} \bar{c}_{L, \mathrm{tvb}}^{n, p}(\hmv{\xi}; \mv{\mu}) \geq 0 \\
&\hspace{50pt} \bar{c}_{L, \mathrm{tvb}}^{n, v}(\hmv{\xi}; \mv{\mu}) \geq 0,
\end{split}
\end{equation}
wherein we introduced $\bar{c}_{L, \mathrm{tvb}}^{n, u}, \bar{c}_{L, \mathrm{tvb}}^{n, p}$ and $\bar{c}_{L, \mathrm{tvb}}^{n, v}$ as the tvb-constraint \eqref{eq:LSPG tvb-constraint} applied to the individual 1D fields $u$, $p$ and $v$.

The reconstruction using tvb-constrained projection is depicted in Figure
\ref{fig:Euler snapshot projection} (green curve). Table
\ref{tab:relative errors Euler snapshot} depicts relative errors as well as violations of the 150\% tvb-constraint obtained by the two investigated projections.  We observe an improvement
in velocity and specific volume field accuracy at an expense of the pressure
field reconstruction accuracy, oscillations in velocity and specific volume field could be reduced. 

\begin{table}[h!]
\centering
\begin{tabular}[h]{c|c|c|c|c|c|c|c|c}
& \multicolumn{4}{c|}{orthogonal projection} & \multicolumn{4}{c}{constrained projection} \\
\hline 
error & velocity & pressure & specific volume & total error & velocity & pressure & specific volume & total error \\
\hline
$\varepsilon_{\mathrm{x}}^{n}$ & 4.9\% & 0.073\% & 12.4\% & 0.019\% & 2.7\% & 6.1\% & 2.3\% & 6.1\% \\
\hline
$\varepsilon_{\mathrm{tvb}}^{n}$ & 73.66 & 0 & 0.18 & - & 0 & 0 & 0 & -
\end{tabular}
\caption{Relative state errors and violation of the tvb-constraint for
	projection of Euler equations solution snapshot on an affine subspace.} 
\label{tab:relative errors Euler snapshot}
\end{table}

\subsection{Burgers' equation}
\label{sec:Burgers' equation}
We demonstrate the performance of the proposed formulation on the source-free
and inviscid Burgers' equation 
\begin{equation}
\label{eq:Burgers' continuos problem}
\frac{\partial u(z, t; \mv{\mu})}{\partial t} + u(z, t; \mv{\mu}) \frac{\partial u(z, t; \mv{\mu})}{\partial z} = 0, \quad \forall (z,t) \in [0, L] \times [0, T]
\end{equation}
with the parameterized initial condition
\begin{equation}
u(0, z; \mv{\mu}) = u_i(z; \mv{\mu}) \equiv \mu_2  \cos\left(\frac{2\pi\mu_1}{100} z\right) + (\mu_2 + 1), \quad \forall z \in \ (0,L]
\end{equation}
and the Dirichlet boundary condition
\begin{equation}
u(0, t; \mv{\mu}) = u_i(0; \mv{\mu}), \quad \forall  t \in [0, T].
\end{equation}
Here, $u(z,t;\mv{\mu})$ denotes the solution of interest, $z \in [0, L]$
denotes the spatial coordinate (with $L=100$ being the right boundary), $t \in
[0, T]$ denotes time (with $T=30$ the final time) and $\mv{\mu}\equiv(\mu_1,
\mu_2)$ denote initial-condition parameters corresponding to the
frequency and the amplitude (and height), respectively. The parametric domain bounds are
given by $\mu_1 \in [0.8; 1.2 ]$, $\mu_2 \in [0.2; 0.6]$.

\subsubsection{Discretization and offline training}
We apply a finite-volume scheme with an upwind flux on a {uniformly spaced grid, yielding a model with
$N=200$ (i.e., 200 control volumes) degrees of freedom. We apply a backward
Euler time discretization with $N_T = 150$ time instances such that $\Delta t
= 0.2$. Figure \ref{fig:burgers numerical solution example} depicts 
the solution at several time steps for a specific parameter instance. 

\begin{figure}[!htb]
  \centering
  \includegraphics[width=0.4\linewidth]{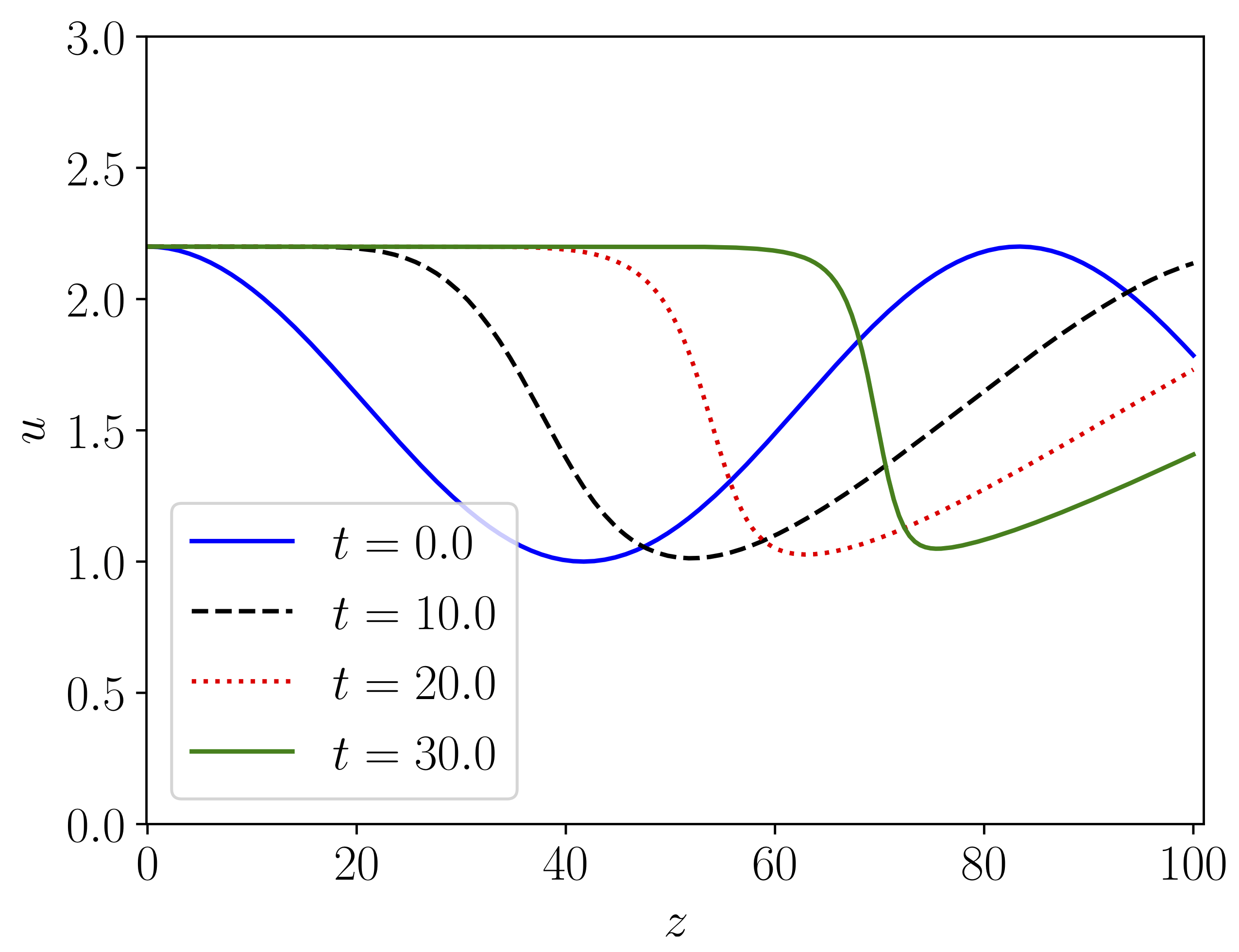}
  \caption{ Numerical solution of Burgers' equation at different time steps.
	The depicted solution corresponds to a parametrization of $\mv{\mu} = (1.2,
	0.6)$.} 
  \label{fig:burgers numerical solution example}
\end{figure}

Training simulations are evaluated on the training set 
\begin{equation}
\label{eq:training set Burgers}
\mathcal{D}_{\mathrm{train}} \equiv \{\mv{\mu}_{\mathrm{train}}^i\}_{i=1}^{n_{\mathrm{train}}}, 
\quad \mathrm{with} \  n_{\mathrm{train}}=4
\end{equation}
resulting from all combination of $\mu_1 \in \{0.8, 1.2\}$ and $\mu_2 \in
\{0.2, 0.6\}$. The reduced-order basis $\mv{\Phi} \in V_{p}(\RR^{200})$ is computed from the first $p$ POD modes of the accumulated snapshot matrix.

\subsubsection{Applied constraints}
 We apply a rsum-constraint of the forms \eqref{eq:Galerkin conservation
constraint}  and \eqref{eq:LSPG conservation constraint} for constrained
Galerkin and LSPG projection, respectively, with a single
equality constraint over the sum of all residual entries, that is $\mv{C} =
[1,1,\hdots,1]$ with $\mv{C} \in \RR^{1 \times 200}$. 
This constraint can be interpreted as global conservation under
certain assumptions, see Ref.~\cite{Carlberg2018280} for a detailed
discussion.
Furthermore, given that the
Burgers' equation under consideration is source-free, the numerical solution of
the FOM is total variation diminishing, such that we enforce an inequality
constraint of the forms \eqref{eq:Galerkin TVD constraint} and \eqref{eq:LSPG
TVD constraint} for constrained
Galerkin and LSPG projection, respectively.

\subsubsection{Results}
All results reported in this section were produced using the functionality of scipy.optimize.minimize (\textit{method}='SLSQP' with parameters \textit{maxiter}=100, \ \textit{ftol}=$10^{-10}$, cf. Section \ref{sec:Nonlinear solution strategies}).
Figures \ref{fig:burgers gp error metrics} and \ref{fig:burgers lspg error
metrics} depict error metrics for the Burgers' flow choosing a reduced
dimension of $p=10$. Every model is evaluated at two online parametric
configurations, namely $\mv{\mu}_1 = (0.9, 0.3)\not\in\mathcal{D}_{\mathrm{train}}$ and $\mv{\mu}_2 = (1.3,
0.7)\not\in\mathcal{D}_{\mathrm{train}}$. Note that $\mv{\mu}_1$ and $\mv{\mu}_2$ correspond to configurations
with different complexity in the sense that $\mv{\mu}_1$ is a setting within
the bounds of the training set $\mathcal{D}_{\mathrm{train}}$ while
$\mv{\mu}_2$ resides outside the training set bounds. 

The global rsum-constraint, which can be found in subfigures (a) and (b)
indicates that with rsum being in use the corresponding error metrics
$\varepsilon^{n}_{G, \mathrm{rsum}}$ and $\varepsilon^{n}_{L, \mathrm{rsum}}$
decrease locally by several orders of magnitude and achieve near machine
precision; this implies that these constraints were feasible.

The violation of the tvd-constraint is depicted in subfigures (c) and (d).
Configuration $\mv{\mu}_1$ fulfills the tvd-constraint for every combination
of constraints, even the simple unconstrained ROM is tvd. This changes for
configuration $\mv{\mu}_2$, for which the unconstrained ROM can not maintain
diminishing total variation over the entire time of simulation. Turning the
tvd-constraint on yields a regain of the desired property, forcing the ROM to
inherit the tvd-property from the FOM. This implies that the tvd-constraint
was indeed feasible. Subfigures (e) and (f) indicate that
the constraints do not have a negative effect on the state error.

Figure \ref{fig:burgers variable basis size} depicts error metrics for
$\mv{\mu}_2$ over a variable reduced-order basis size $p$. A refinement in $p$
results in decreasing error metrics for the unconstrained models. However, the
rsum- as well as the tvd-constraint yield better accuracy in the corresponding
error metrics over the entire range, especially for small basis dimensions.

\subsection{Euler equations}
\label{sec:Euler equations}
We now present an ideal gas flow described by the 1D Euler equations
\begin{align}
\frac{\partial u(z, t; \mv{\mu})}{\partial t} + u(z, t; \mv{\mu}) \frac{\partial u(z, t; \mv{\mu})}{\partial z} + v(z, t; \mv{\mu}) \frac{\partial p(z, t; \mv{\mu})}{\partial z} &= 0, 
\quad \forall (z, t) \in [0, L] \times [0,T], \\
\frac{\partial p(z, t; \mv{\mu})}{\partial t} + u(z, t; \mv{\mu}) \frac{\partial p(z, t; \mv{\mu})}{\partial z} + \gamma p(z, t; \mv{\mu}) \frac{\partial u(z, t; \mv{\mu})}{\partial z} &= 0, 
\quad \forall (z, t) \in [0, L] \times [0,T], \\
\frac{\partial v(z, t; \mv{\mu})}{\partial t} + u(z, t; \mv{\mu}) \frac{\partial v(z, t; \mv{\mu})}{\partial z} - v(z, t; \mv{\mu}) \frac{\partial u(z, t; \mv{\mu})}{\partial z} &= 0, 
\quad \forall (z, t) \in [0, L] \times [0,T],
\end{align}
where the solution is composed of the velocity field $u(z, t;
\mv{\mu})$, the pressure field $p(z, t; \mv{\mu})$, and the mass-specific volume field
$v(z, t; \mv{\mu})$; here, $\gamma$ denotes the isentropic exponent. Parametrization
in terms of the 2D parameter vector $\mv{\mu} \equiv (\mu_1, \mu_2)$ is
introduced within the initial condition as well as the boundary condition. The
initial condition reads 
\begin{align}
u(z, 0; \mv{\mu}) = u_0, \quad 
p(z, 0; \mv{\mu}) = p_0, \quad 
v(z, 0; \mv{\mu}) = \frac{1}{\mu_1}, \quad  \forall z \in (0, L], 
\end{align}
and we denote the boundary condition by 
\begin{align}
u(0, t; \mv{\mu}) = u_{\mathrm{dbc}} = u_0, \quad 
p(0, t; \mv{\mu}) = p_{\mathrm{dbc}} = p_0 \mu_2, \quad 
v(0, t; \mv{\mu}) = v_{\mathrm{dbc}} = \frac{1}{\mu_1 \mu_2}, \quad  \forall t \in [0, T].
\end{align}
The parametrization can be physically interpreted as 
\begin{align}
\mu_1 &= \rho_0,\qquad
\mu_2 = \frac{p_{\mathrm{dbc}}}{p_0} = \frac{\rho_{\mathrm{dbc}}}{\rho_0}. \label{eq:Euler flow mu_2}
\end{align}
That is, $\mu_1$ corresponds to the initial density of the fluid, while
$\mu_2$ corresponds to the ratio of the pressure boundary condition and the initial
pressure value and at the same time the ratio of the density boundary
condition and the initial density. Consequently, the described setup can be
interpreted as a 1D model of a tube, which is continuously traversed by an
ideal gas, while at a specific instance in time ($t=0$), a sudden change in the
fluid properties appears at the inlet of the tube. 

The constants introduced are set to $\gamma = 1.4$, $u_0 = 400 \frac{\mathrm{m}}{s}$, $p_0 = 101000 \mathrm{Pa}$, $L = 1.25 \mathrm{m}$ and $T = 0.001 \mathrm{s}$, and the parametric domain bounds are given by $\mu_1 \in [1.1 \frac{\mathrm{kg}}{\mathrm{m}^3}; 1.6 \frac{\mathrm{kg}}{\mathrm{m}^3}]$, $\mu_2 \in [1.1; 1.4]$.
Given that the speed of sound is $c = \sqrt{\frac{p_0}{\rho_0} \gamma} < u_0$ (recall that $\rho_0 = \mu_1$), the flow under consideration is classified as supersonic.

\begin{figure}[!htb]
\centering
\includegraphics[width=0.8\linewidth]{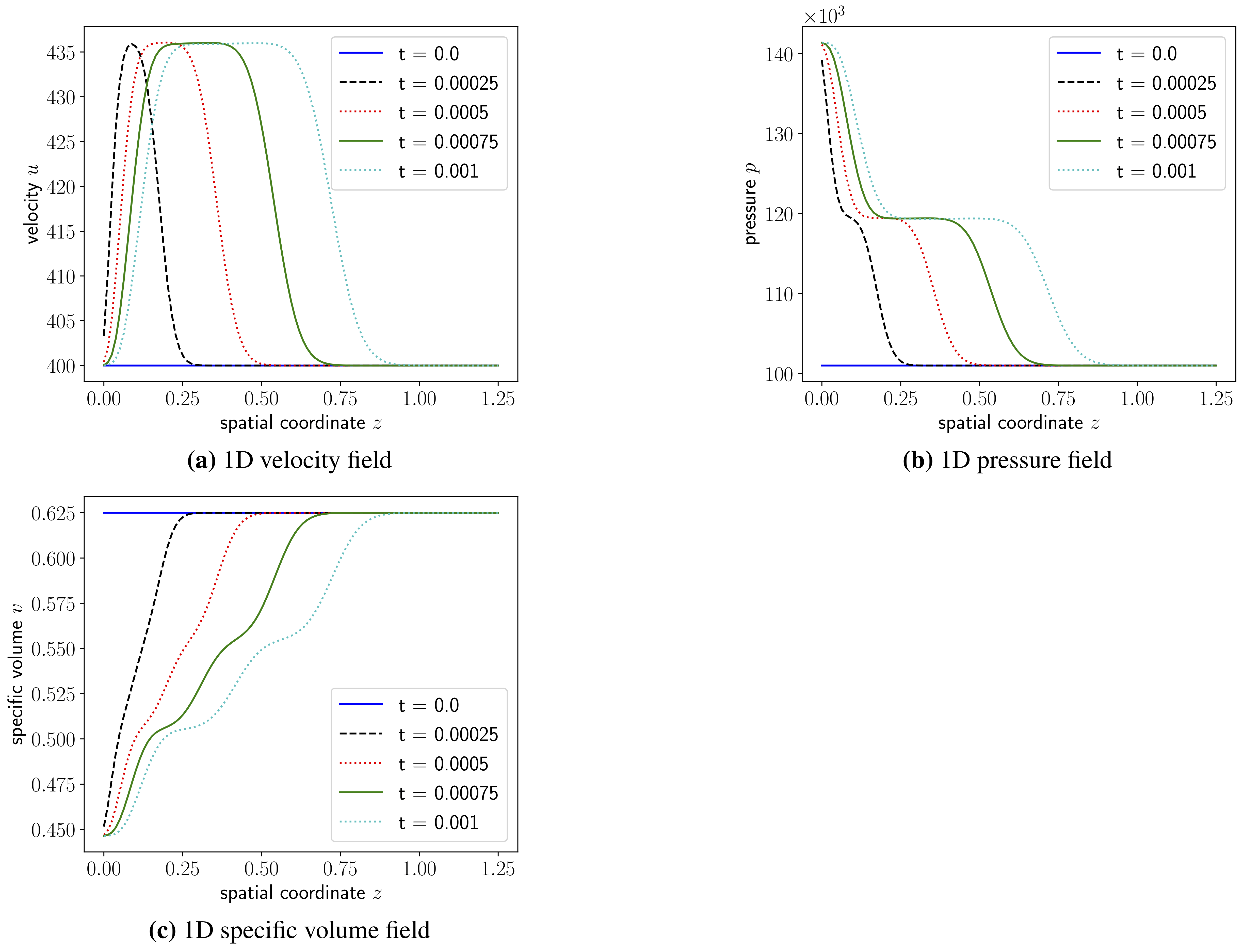}
\caption{Numerical solution of Euler equations at different time steps. The depicted solution corresponds to a parametrization of $\mv{\mu} = (1.6, 1.4)$.}
\label{fig:Euler numerical solution}
\end{figure}

\subsubsection{Discretization and offline training}
We apply a finite-difference spatial discretization scheme on a uniform
grid with backward
finite differences and $N = 300$ (i.e., 100 degrees of freedom for $u, p, v$
each) and $N_T$ time steps ($N_T = 200$ and $N_T = 600$ will be considered further down below). Time stepping is performed with the explicit
Euler method, see Figure \ref{fig:Euler numerical solution} for an exemplary
solution at several time steps for a specific parameter instance. 

Training simulations are evaluated on the training set 
\begin{equation}
\label{eq:training set Euler}
\mathcal{D}_{\mathrm{train}} \equiv \{\mv{\mu}_{\mathrm{train}}^i\}_{i=1}^{n_{\mathrm{train}}}, 
\quad \mathrm{with} \  n_{\mathrm{train}}=4
\end{equation}
resulting from all combination of $\mu_1 \in \{1.1, 1.6\}$ and $\mu_2 \in \{1.1, 1.4\}$. A reduced-order basis $\mv{\Phi} \in V_{p}(\RR^{300})$ is computed from the first $p$ POD modes of the training simulation snapshots.

\begin{remark}
Unconstrained Galerkin and LSPG projection ROMs are equivalent for time discretization by an explicit linear multistep scheme, see Ref.~\cite{carlberg2017galerkin}. As a consequence, we depict only one curve for both unconstrained projections in \ref{sec:appendix Euler equations}.
\end{remark}

\subsubsection{Applied constraints}
We apply a quantity-specific tvb-constraint of the forms \eqref{eq:Galerkin tvb-constraint} and \eqref{eq:LSPG tvb-constraint}. By quantity-specific we mean that total variation is bounded for each quantity ($u$, $p$ and $v$) independently, which results in three inequality constraints. Three total variation bounds (one for each quantity) are computed as 120 \% of the quantity-specific maximum total variation values of all training simulation snapshots.

\begin{remark}
Estimating a total variation bound from training simulation snapshots can be
	seen as a consistent extension of the snapshot collection method for the
	construction of a low-dimensional solution subspace, which corresponds to
	kinematically constraining the space of all admissible solution states. In
	this context, a total variation bound computed from solution snapshots
	simply corresponds to further kinematically constraining the state.  
\end{remark}  

\subsubsection{Results}
All results reported in this section were produced using the functionality of scipy.optimize.minimize (\textit{method}='trust-constr' with parameters \textit{maxiter}=250, \textit{gtol}=$10^{-6}$, \textit{xtol}=$10^{-6}$, \textit{barrier\textunderscore tol}=$10^{-6}$ for Galerkin projection and \textit{method}='SLSQP' with parameters \textit{maxiter}=250, \textit{ftol}=$10^{-6}$ for LSPG projection, cf. Section \ref{sec:Nonlinear solution strategies}).
Figures \ref{fig:Euler problem Galerkin projection state errors} and
\ref{fig:Euler problem Galerkin projection total variations} depict results
collected from constrained and unconstrained Galerkin and LSPG projections choosing a
reduced dimension of $p=20$. Plots in the left column are computed from 200
time steps, while plots in the right column are computed from a finer
discretization in time using 600 time steps. Figure \ref{fig:Euler problem
Galerkin projection state errors} illustrates relative state errors
\eqref{eq:relative state space error} for the three quantities (velocity
$\varepsilon^{n}_{x,u}$, pressure $\varepsilon^{n}_{x,p}$, specific volume
$\varepsilon^{n}_{x,v}$) involved in the solution of the Euler equations. The
presence of the tvb-constraint improves the state errors. On closer
inspection, the plots show an overlap of constrained and unconstrained curves
up to a certain point in time, at which the curves split. This observation can
be explained by inspecting  Figure \ref{fig:Euler problem Galerkin
projection total variations}, which depicts quantity-specific total variations
as well as the applied total variation bound. As a result, the curves in
Figure \ref{fig:Euler problem Galerkin projection state errors} overlap until
the tvb-constraint is active. When the unconstrained curve crosses the total
variation bound, state errors start to separate. 

Also the total variations in Figure \ref{fig:Euler problem Galerkin projection
total variations} overlap initially and split, when the tvb-constraint becomes active. However, tvb-constrained curves (especially pronounced for velocity
and pressure) still run above the total variation bound for the Galerkin projection, such that the
tvb-constraint \eqref{eq:tvb-constraint} is not fulfilled. We reference Remark
\ref{remark:kinematic inequality constraint for Galerkin projection} for the
explanation of this behavior, highlighting that the constraint violation at
hand is a consequence of time discretization. Using a finer discretization in
time (600 instead of 200 time steps) results in lower constraint violation for the Galerkin projection,
which can be seen from right column plots of Figure \ref{fig:Euler problem
Galerkin projection total variations}. For the LSPG projection, the applied total variation bound is strictly respected for 200 as well as 600 time steps.

Figure \ref{fig:Galerkin projection mean global error metrics for Euler problem} depicts mean global error metrics. Discussing the unconstrained ROM first, we observe an initial increase of the state error for all quantities with a peak at $p=20$, followed by a sudden decrease with increasing reduced-order basis size until $p=30$. The tvb-constraint violation plots explain the peak at $p=20$, showing that the unconstrained ROM solution becomes oscillatory. A further increase in $p$ until $p=30$ leads to a rather accurate ROM, which reflects in the state space error as well as the tvb-constraint violation plots. For further increase in $p$, the unconstrained ROM again becomes oscillatory at $p=35$ and eventually unstable at $p=40$. 

In contrast, both tvb-constrained Galerkin and LSPG projection ROMs show only slight (Galerkin projection) or no (LSPG projection) violation of the tvb-constraint over all evaluated reduced-order basis sizes and also achieve a stable solution, when the unconstrained ROM becomes unstable ($p=40$).

We observe that state errors for some $p$-values are slightly worse with tvb-constraint activated, while the tvb-constraint is almost fulfilled by the unconstrained curve ($p=25$ and $p=30$). The conclusion from the current example is that the tvb-constraint is helpful when the unconstrained ROM severely violates that property, while it can become harmful (for the state error) when the unconstrained ROM is sufficiently accurate by itself.

\subsection{Diffusion equation}
\label{sec:Diffusion equation}
We solve a nonlinear, non-homogeneous diffusion equation of the shape
\begin{equation}
\rho c \frac{\partial \theta(t, y, z; \mv{\mu})}{\partial t} - \nabla \cdot (\lambda(\theta) \nabla \theta(t, y, z; \mv{\mu})) = s(t, y, z; \mv{\mu}), \quad \forall (t, y, z) \in [0, T] \times [0,L] \times [0,L]
\end{equation}
with the initial condition
\begin{equation}
\theta(0, y, z; \mv{\mu}) = 300, \quad \forall (y,z) \in [0,L] \times [0,L],
\end{equation}
adiabatic boundary conditions
\begin{equation}
\label{eq:adiabatic boundary condition diffusion problem}
\nabla \theta(t, y, z; \mv{\mu}) \cdot \mv{n} = 0, \quad \forall (t,z,y) \in [0,T] \times \{0,L\} \times \{0,L\},
\end{equation}
a parameterized source 
\begin{equation}
s(t, y, z; \mv{\mu}) = 
\underbrace{\mu_1 \delta(y - (0.5 + \mu_4 \sin(2 \pi \mu_3 t)) \delta(z - 0.2)}_{s_1} + \underbrace{\mu_2 \delta(y - (0.5 - \mu_4 \sin(2 \pi \mu_3 t)) \delta(z - 0.5)}_{s_2},
\end{equation}
and a temperature dependent heat conductivity given by 
\begin{equation}
\lambda(\theta) = (\theta - 300) + 50.
\end{equation}
Therein we introduced the area density $\rho = 8000 \ \mathrm{kg} / \mathrm{m}^2$, mass-specific heat capacity $c = 500 \ \mathrm{J} /( \mathrm{kg\  K})$, temperature $\theta \ [\mathrm{K}]$, thermal conductivity $\lambda [\mathrm{W} / (\mathrm{m \ K})]$, the time $t \in [0, T]$ (with $T = 10000 \ \mathrm{s}$ being the end time) and spatial coordinates $(y,z) \in [0,L] \times [0,L]$ (with $L = 1 \ \mathrm{m}$ describing the domain size). The parameterized source $s$ consists of two dirac delta $\delta$ pulses $s_1$ and $s_2$, which are moving in $y$-direction in time. Thereby $\mu_1 \ [\mathrm{W} / \mathrm{m}^2]$ and $\mu_2 \ [\mathrm{W} / \mathrm{m}^2]$ are the source strengths, $\mu_3 \ [1/\mathrm{s}]$ is the moving frequency and $\mu_4 \ [\mathrm{m}]$ the amplitude. 

\subsubsection{Discretization and offline training}
We apply a central finite difference scheme for spatial discretization on a uniform grid
yielding $N = 1089$ degrees of freedom (33 in each coordinate axis), and employ the implicit
Euler time discretization method with $N_T = 100$ time steps. Figure
\ref{fig:diffusion problem solution} depicts an exemplary solution at 
parameter instance $\mv{\mu} = (-10^6, 10^6, 10^{-4}, 0.2)$ including a source
and a sink (negative source strength).

Training simulations are evaluated on the training set 
\begin{equation}
\label{eq:training set diffusion}
\mathcal{D}_{\mathrm{train}} \equiv \{\mv{\mu}_{\mathrm{train}}^i\}_{i=1}^{n_{\mathrm{train}}}, 
\quad \mathrm{with} \  n_{\mathrm{train}}=8
\end{equation}
resulting from all combinations of $(\mu_1, \mu_2) \in \{(-1.1 \times 10^6,
0.9 \times 10^{-6}), (-10^6, 10^{-6}) \}, \mu_3 \in \{5 \times 10^{-5}, 15
\times 10^{-5}\}, \mu_4 \in \{0.1, 0.3\}$ and the reduced-order basis
$\mv{\Phi} \in 
V_{p}(\RR^{1089 })$ is computed from the first $p$ POD modes
of the accumulated snapshot matrix.

\begin{figure}[!htb]
\centering
\includegraphics[width=0.8\linewidth]{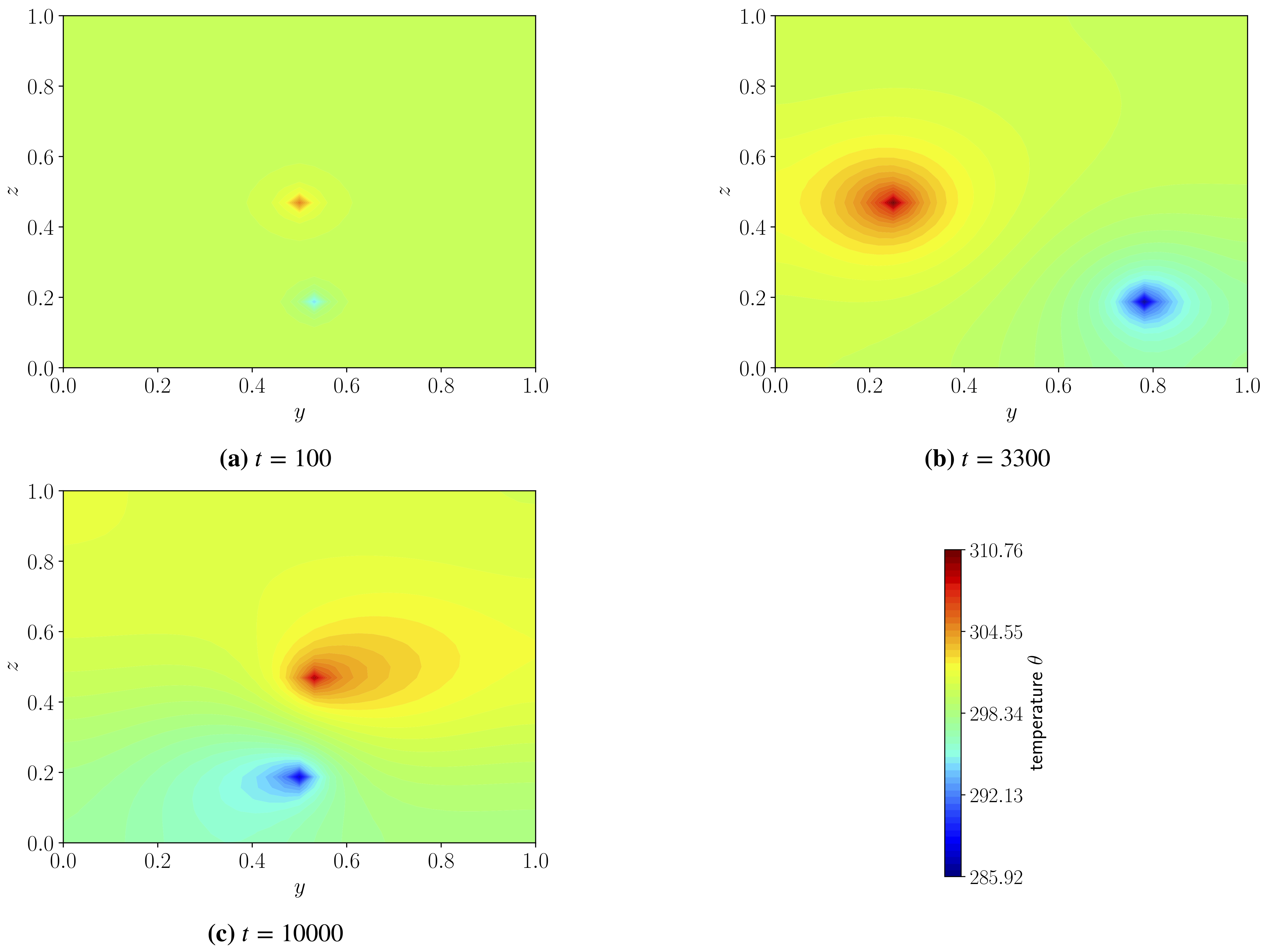}
\caption{Numerical solution of nonlinear diffusion equation at different time
	steps. The depicted solution corresponds to a parametrization of $\mv{\mu} =
	(-10^6, 10^6, 10^{-4}, 0.2)$.}
\label{fig:diffusion problem solution}
\end{figure}

\subsubsection{Applied constraints}
We investigate the performance of the proposed reduced-order models for the exemplary parametrization 
\begin{equation}
\label{eq:diffusion parametrization}
	\mv{\mu} = (-10^{6}, 10^6, 10^{-4}, 0.2)
\end{equation}
applying a rsum-constraint of the forms \eqref{eq:Galerkin conservation constraint} and \eqref{eq:LSPG conservation constraint} and an ec-constraint of the forms \eqref{eq:Galerkin projection ec-constraint} and \eqref{eq:LSPG projection ec-constraint}. The rsum-constraint is set to two equality constraints overlapping in 100 elements, that is 
\begin{equation}
\mv{C} = 
\left[
\begin{array}{l}
1 \ \ \hdots \\
0\hdots0
\end{array}
\right.
\hspace{-5pt}
\underbrace{
\begin{array}{l}
1\hdots1 \\
1\hdots1
\end{array}}_{100}
\left.
\hspace{-5pt}
\begin{array}{l}
0\hdots0 \\
\hdots \ \ 1
\end{array}
\right] 
\in \RR^{2 \times 1089}.
\end{equation} 
The ec-constraint resolves to a simple form in case of the investigated computational example. The global energy of the computational domain in the continuous setting results from 
\begin{equation}
e = \int_0^1 \int_0^1 \rho c \theta(t, y, z; \mv{\mu})\ dydz = \rho c \int_0^1 \int_0^1 \theta(t, y, z; \mv{\mu})\ dydz.
\label{eq:global energy content continuous temperature}
\end{equation}  
Assuming $\mv{x}(t; \mv{\mu})$ to be a numerical approximation of the (spatially uniform) discretized temperature field $\theta(t, y, z; \mv{\mu})$, global energy \eqref{eq:global energy content continuous temperature} can be expressed as
\begin{equation}
e = E(\mv{x}) = \alpha \rho c \mv{1}^T \mv{x},
\end{equation}
wherein $\alpha$ is a proportionality constant depending on the spatial discretization (in our case $\alpha = \frac{1}{N}\mathrm{m}^2 = \frac{1}{1089} \mathrm{m}^2$) and $\mv{1}$ is a vector of ones. Additionally, source and sink in \eqref{eq:diffusion parametrization} are of the same absolute strength, which (recalling the adiabatic boundary condition \eqref{eq:adiabatic boundary condition diffusion problem}) results in a vanishing energy intake $S(t;\mv{\mu})$ such that energy conservation \eqref{eq:definition ec-constraint} can be expressed as 
\begin{equation}
\alpha \rho c \mv{1}^T \dmv{x} = 0
\end{equation} 
or simply 
\begin{equation}
\mv{1}^T \dmv{x} = 0.
\end{equation}

\subsubsection{Results}
All results reported in this section were produced using the functionality of scipy.optimize.minimize (\textit{method}='SLSQP' with parameters \textit{maxiter}=100, \textit{ftol}=$10^{-10}$, cf. Section \ref{sec:Nonlinear solution strategies}).
Figure \ref{fig:Galerkin projection diffusion problem} depicts the violation
of the rsum-constraint \eqref{eq:violation of the Gelerkin projection
rsum-constraint}, \eqref{eq:violation of the LSPG projection rsum-constraint},
the deviation of system energy from its initial state \eqref{eq:deviation of
system energy} and the relative state error \eqref{eq:relative state space
error} for both Galerkin projection and LSPG projection at the parametric
configuration \eqref{eq:diffusion parametrization} and a reduced dimension of $p=10$. 
Note that in case of the
present example, energy conservation corresponds to a constant system energy,
such that energy deviation from the initial state can be interpreted as a
violation of conservation of energy. As a result, the rsum-constraint
decreases the corresponding error metric by several orders of magnitude to
nearly zero, implying feasibility of the rsum-constraint. 
Similarly, the error in energy conservation drops to nearly zero
whenever the ec-constraint is activated, also implying feasibility of this
constraint.
However, both of these constraints have negligible effect on the state error
in this particular case.

Figure \ref{fig:Galerkin projection mean global error metrics} reports global
error metrics as a function of ROM dimension $p$. In all cases, the state
error decreases monotonically with increased basis dimension. For any reduced
dimension, enforcing the appropriate constraint has a more substantial effect
on satisfying the rsum-constraint and ec-constraint than simply increasing the
basis dimension. For constrained Galerkin projection, we
observe an increased state error for $p=5$ with rsum- and ec-constraint
activated. Obviously the three constraints (two for rsum, one for ec) restrict the
five-dimensional solution space in a way that shows a negative impact on the
state error. However, we emphasize that this effect vanishes as the ROM
dimension increases.

\section{Conclusion}
\label{sec:Conclusion}
This work proposed a novel constrained-optimization formulation model
reduction of dynamical systems. In particular, the work proposed both
constrained Galerkin and LSPG projection methods that can handle arbitrary
kinematic and dynamic constraints of both equality and inequality type.
This general framework can be used as a template to enforce particular
problem-specific physical properties. This work presented several such
examples, including a zero ``sum-of-residual-entries'' constraint (which can
be used to enforce conservation for finite-volume
models \cite{Carlberg2018280}, a total variation
diminishing constraint, a total variation bounding constraint, and an
energy-conservation constraint. In three numerical examples, we demonstrated
that the proposed framework can indeed be applied in a flexible way to enforce
these desired constraints.

Future work will investigate applying the proposed formulation to large-scale
dynamical-system models, as well as the realization of computational speedup,
which will require including hyper-reduction of the objective and constraints
as needed (as in Ref.~\cite{Carlberg2018280}), and may also entail developing
problem-specific optimizers (e.g., as demonstrated in Refs.~\cite{Zimmermann2014255,zahr2020implicit,corrigan2019moving,Cao2019679}). In addition, we will pursue the extension of the
proposed method to nonlinear trial manifolds as described in Ref.~\cite{lee2018model}; this is straightforward, as it amounts to
replacing the state and velocity kinematic approximations in equations
\eqref{eq:low-dimensional continuous state
representation}--\eqref{eq:low-dimensional velocity representation} with the
non-linear trial manifold counterparts
\begin{align}
	\tmv{x}(t, \mv{\mu}) &\coloneqq \mv{x}_{\mathrm{ref}}(\mv{\mu}) + \mv{g}(
	\hmv{x}(t, \mv{\mu})),
\label{eq:low-dimensional continuous state representation non} \\
	\dot{\tmv{x}}(t, \mv{\mu}) &= \nabla\mv{g}({\hmv{x}}(t, \mv{\mu})) \dot{\hmv{x}}(t, \mv{\mu}),
	\label{eq:low-dimensional velocity representation non} 
\end{align}
with $\mv{g}:\RR^p\rightarrow\RR^N$ a continuously differentiable nonlinear
operator (e.g., computed via an autoencoder).

The generality of the proposed formulation yields both strengths and
weaknesses. Its generality enables it to be used as a template for enforcing a
wide variety of physical constraints; however, efficient solvability of the
resulting specific ROM formulation remains the
responsibility of the end user and may be viewed as a weakness. In our numerical
experiments, we identified two situations that can potentially lead to increased
state errors if the ROM is constrained. The first situation might occur when
the unconstrained ROM is already highly accurate and fulfills the constraints
with reasonable accuracy without explicit enforcement. The second situation
may occur
when the ROM is very low-dimensional (relative to the number of constraints),
such that the constraints significantly reduce the number of effective degrees
of freedom relative to the unconstrained case. In the case of conservation
constraints, we refer the reader to Ref.~\cite{Carlberg2018280}, which
performed a detailed study of the effect of different errors on the number of
constraints and subspace dimension.

\section*{Acknowledgments}
The authors thank Danielle Maddix, Irina Tezaur, Remmelt Ammerlaan, Anran Lu,
Yiwen Guo, and
Kexin Yu for discussions
on this work and related early investigations.
This work was sponsored by Sandia's Advanced Simulation and Computing (ASC)
Verification and Validation (V\&V) Program and by the
Air Force Office of Scientific Research (AFOSR) under award number
FA9550-20-1-0236 (MZ). This paper
describes objective technical results and analysis. Any subjective views or
opinions that might be expressed in the paper do not necessarily represent the
views of the U.S. Department of Energy or the United States Government. Sandia
National Laboratories is a multimission laboratory managed and operated by
National Technology \& Engineering Solutions of Sandia, LLC, a wholly owned
subsidiary of Honeywell International Inc., for the U.S. Department of
Energy’s National Nuclear Security Administration under contract DE-NA0003525.

\clearpage
\bibliography{mybibfile}
\bibliographystyle{siam}

\appendix

\newpage
\pagebreak
\clearpage

\begin{figure}[!htb]
\section{Burgers' equation}
\centering
\includegraphics[width=1.\linewidth]{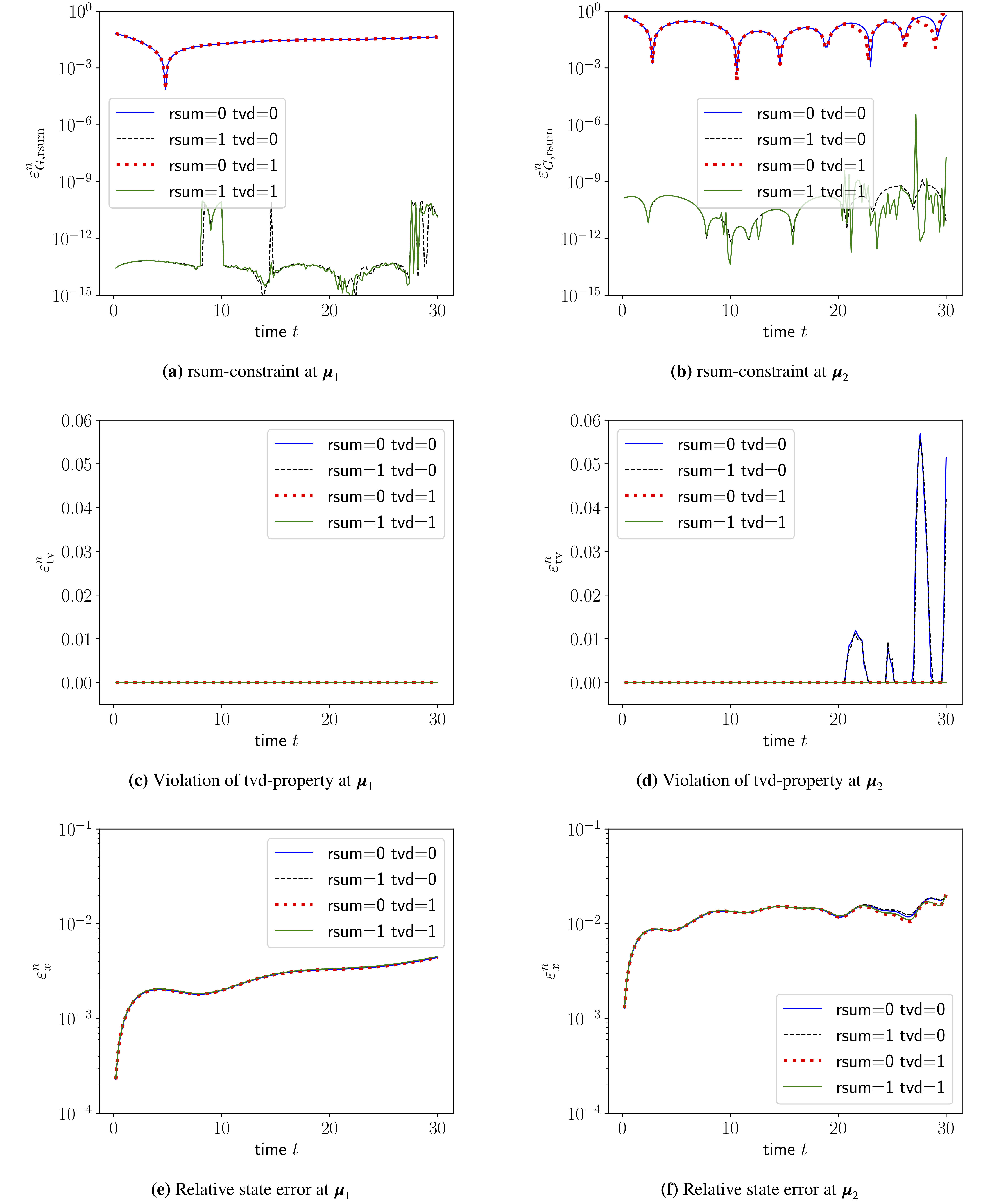} 
\caption{1D Burgers' equation. Error metrics for the Galerkin projection (10 degrees of freedom) evaluated at $\mv{\mu}_1 = (0.9, 0.3)$ and $\mv{\mu}_2 = (1.3, 0.7)$. rsum = 0/1 and tvd = 0/1 denote the sum-of-residual-entries constraint and total variation diminishing constraint being \textit{not in use} / \textit{in use}.}
\label{fig:burgers gp error metrics}
\end{figure}

\begin{figure}[!htb]
\centering
\includegraphics[width=1.\linewidth]{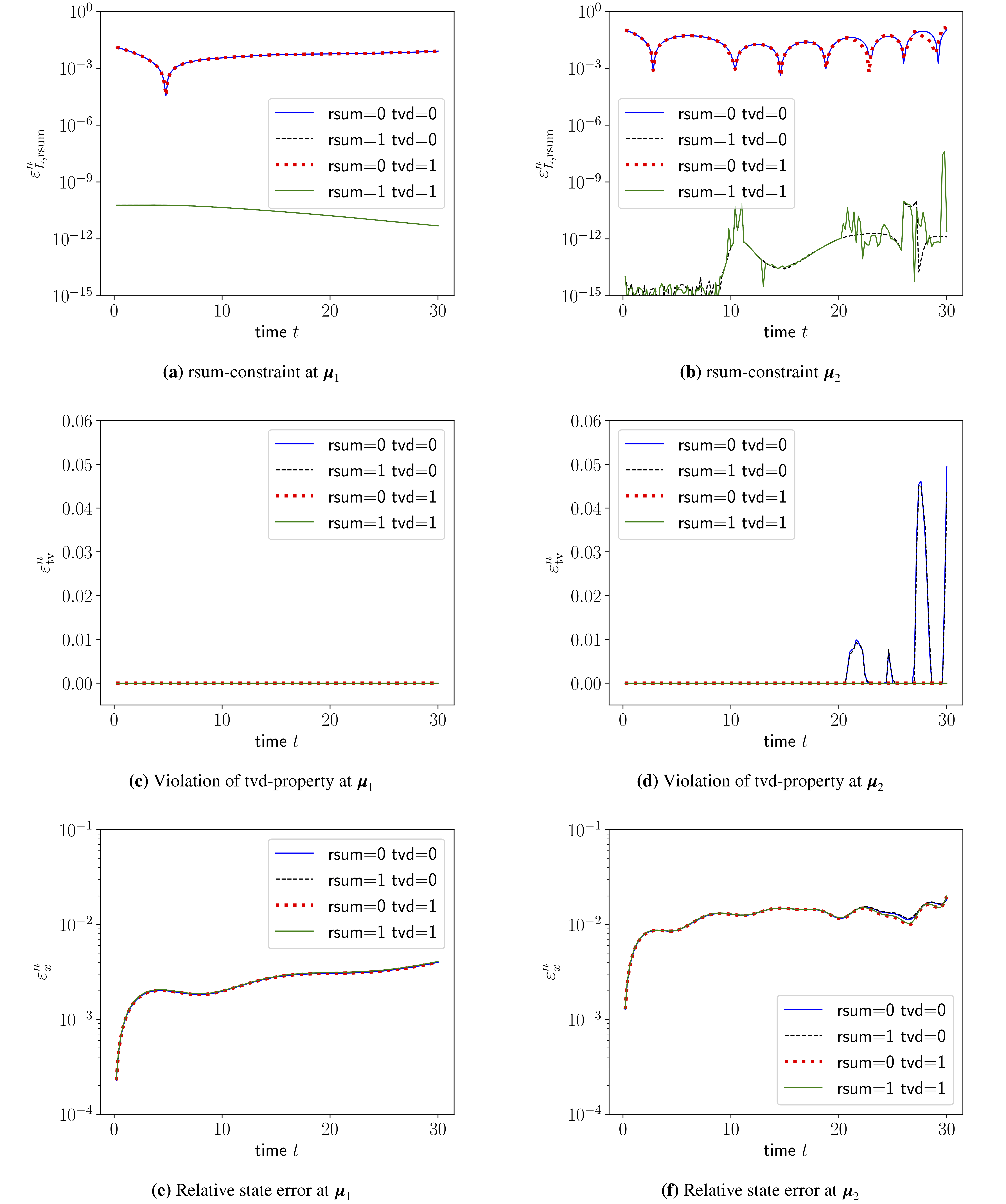} 
\caption{1D Burgers' equation. Error metrics for the LSPG projection (10 degrees of freedom) evaluated at $\mv{\mu}_1 = (0.9, 0.3)$ and $\mv{\mu}_2 = (1.3, 0.7)$. rsum = 0/1 and tvd = 0/1 denote the sum-of-residual-entries constraint and total variation diminishing constraint being \textit{not in use} / \textit{in use}.}
\label{fig:burgers lspg error metrics}
\end{figure}

\begin{figure}[!htb]
\centering
\includegraphics[width=1.\linewidth]{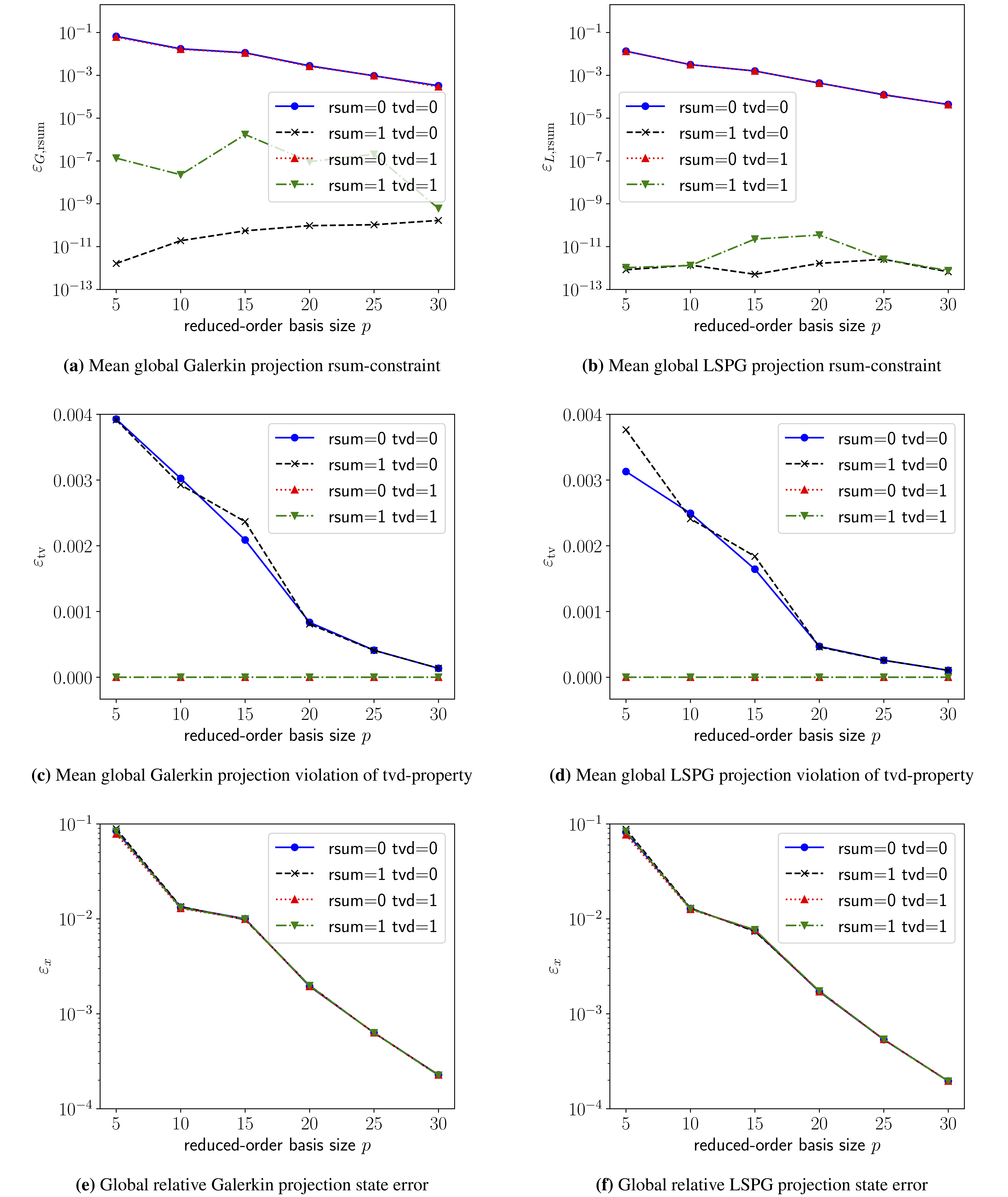} 
\caption{1D Burgers' equation. Galerkin projection (left column) and LSPG projection (right column) global error metrics over variable reduced-order basis size evaluated at $\mv{\mu}_2 = (1.3, 0.7)$. rsum = 0/1 and tvd = 0/1 denote the sum-of-residual-entries constraint and total variation diminishing constraint being \textit{not in use} / \textit{in use}.}
\label{fig:burgers variable basis size}
\end{figure}

\newpage
\pagebreak

\begin{figure}[!htb]
\section{Euler equations}
\label{sec:appendix Euler equations}
\centering
\includegraphics[width=1.\linewidth]{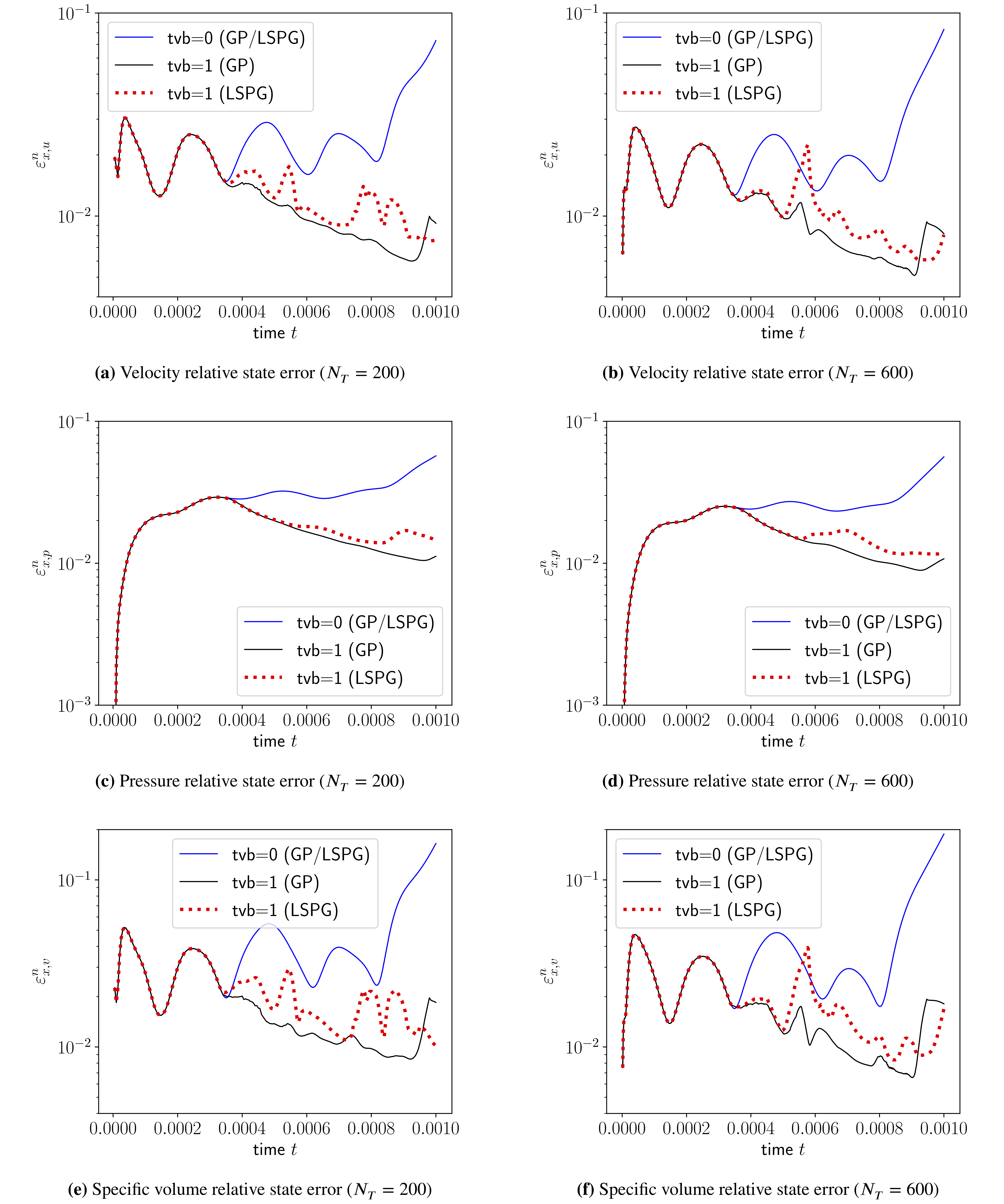} 
\caption{1D Euler equations. State errors for Galerkin projection (20 degrees of freedom) evaluated at $\mv{\mu} = (1.25, 1.5)$. Left column plots were produced with 200 time steps and right column plots with 600 time steps. tvb = 0/1 denotes the total variation bounding constraint being \textit{not in use} / \textit{in use}. Unconstrained GP and LSPG are equivalent due to explicit discretization in time.}
\label{fig:Euler problem Galerkin projection state errors}
\end{figure}

\begin{figure}[!htb]
\centering
\includegraphics[width=1.\linewidth]{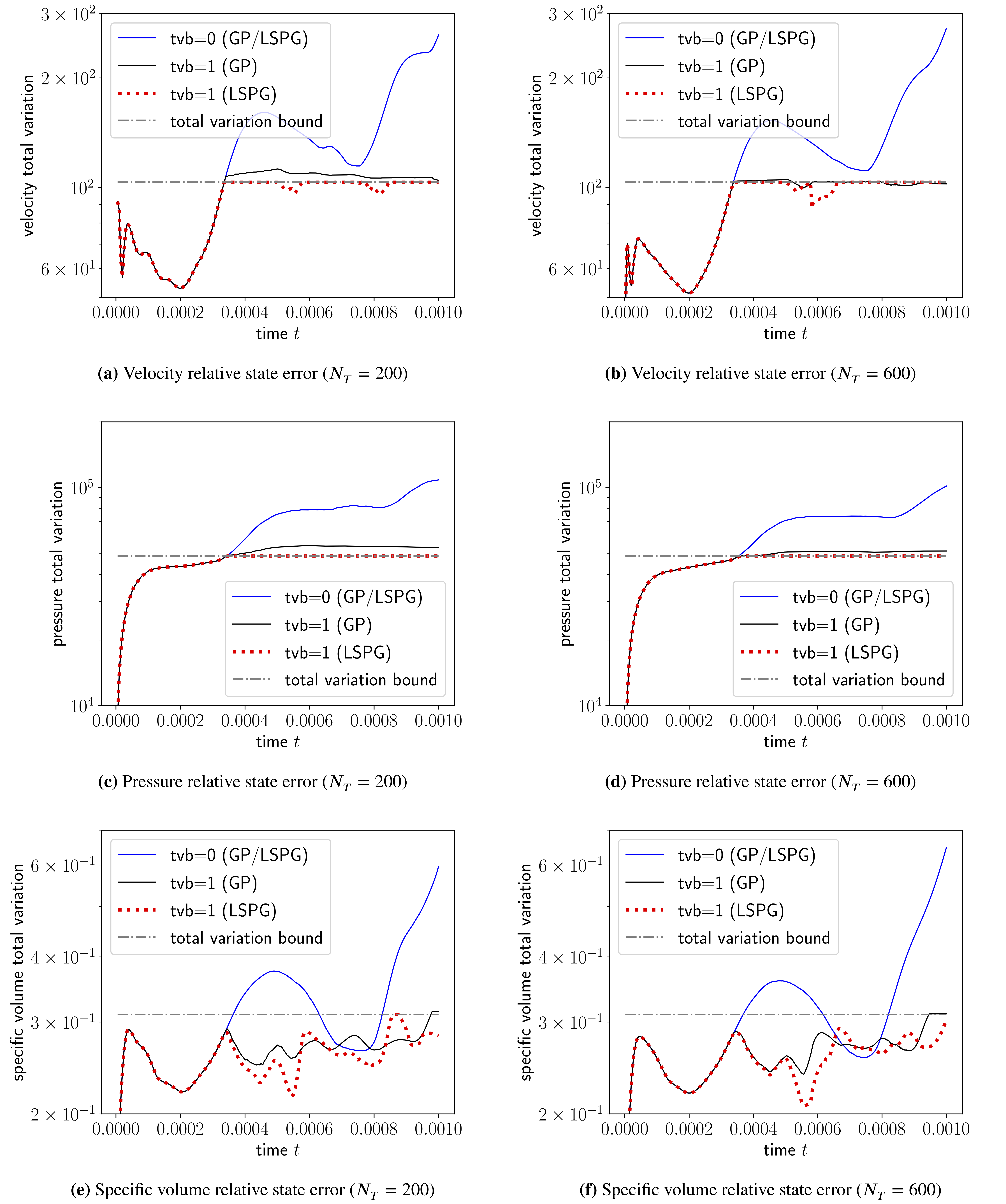} 
\caption{1D Euler equations. State total variations for Galerkin projection (GP) and LSPG projection (LSPG) (20 degrees of freedom) evaluated at $\mv{\mu} = (1.25, 1.5)$. Left column plots were produced with 200 time steps and right column plots with 600 time steps. tvb = 0/1 denotes the total variation bounding constraint being \textit{not in use} / \textit{in use}. Unconstrained GP and LSPG are equivalent due to explicit discretization in time.}
\label{fig:Euler problem Galerkin projection total variations}
\end{figure}

\begin{figure}[!htb]
\centering
\includegraphics[width=1.\linewidth]{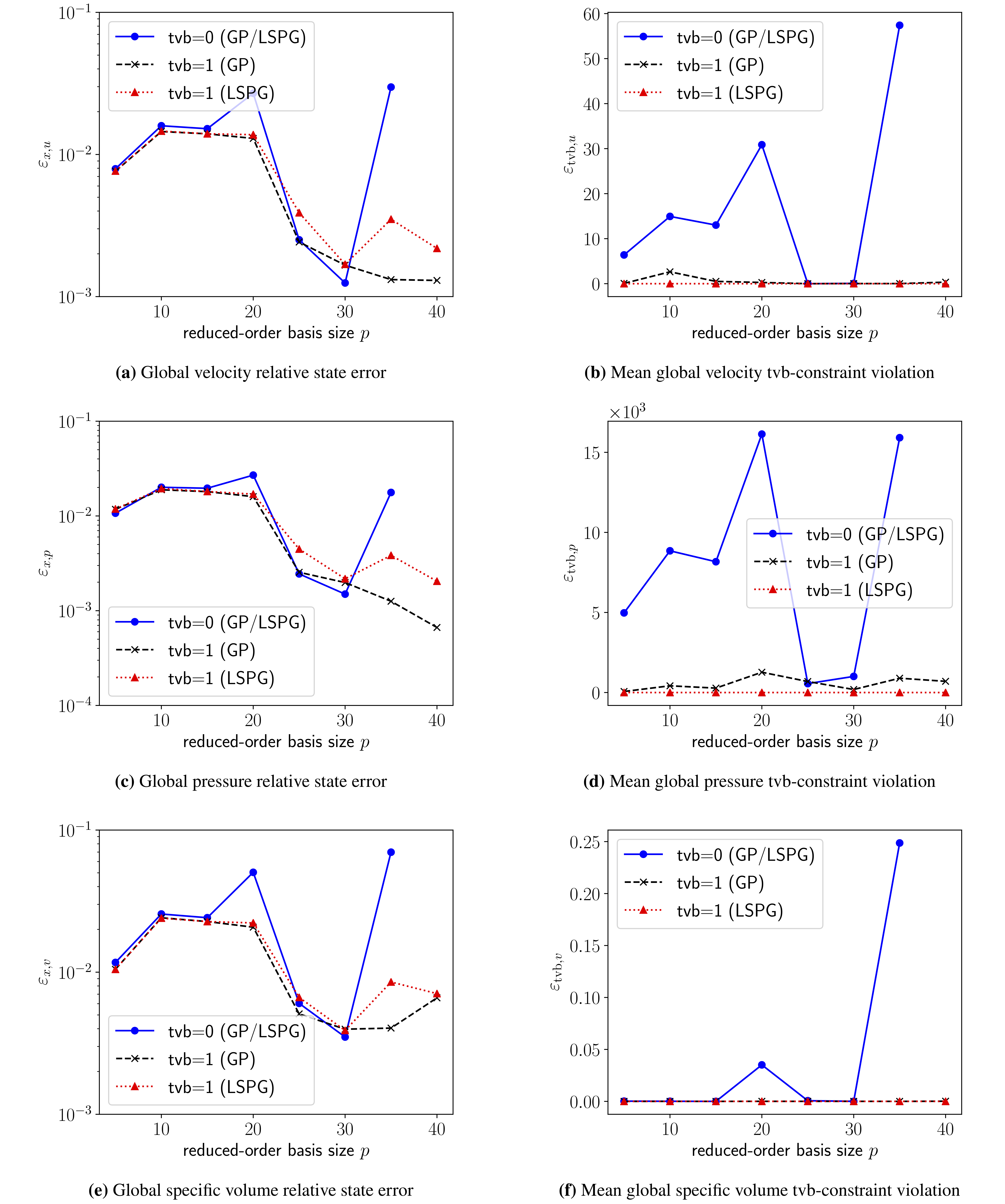} 
\caption{1D Euler equations. Galerkin projection (GP) and LSPG projection (LSPG) global state errors (left column) and mean global total variation bounding constraint violation (right column) at 600 time steps. tvb = 0/1 denotes the total variation bounding constraint being \textit{not in use} / \textit{in use}. The unconstrained projection ROM (tvb=0) was unstable for $p=40$. Unconstrained GP and LSPG are equivalent due to explicit discretization in time.}
\label{fig:Galerkin projection mean global error metrics for Euler problem}
\end{figure}

\pagebreak
\newpage

\begin{figure}[!htb]
\section{Diffusion equation}
\centering
\includegraphics[width=1.\linewidth]{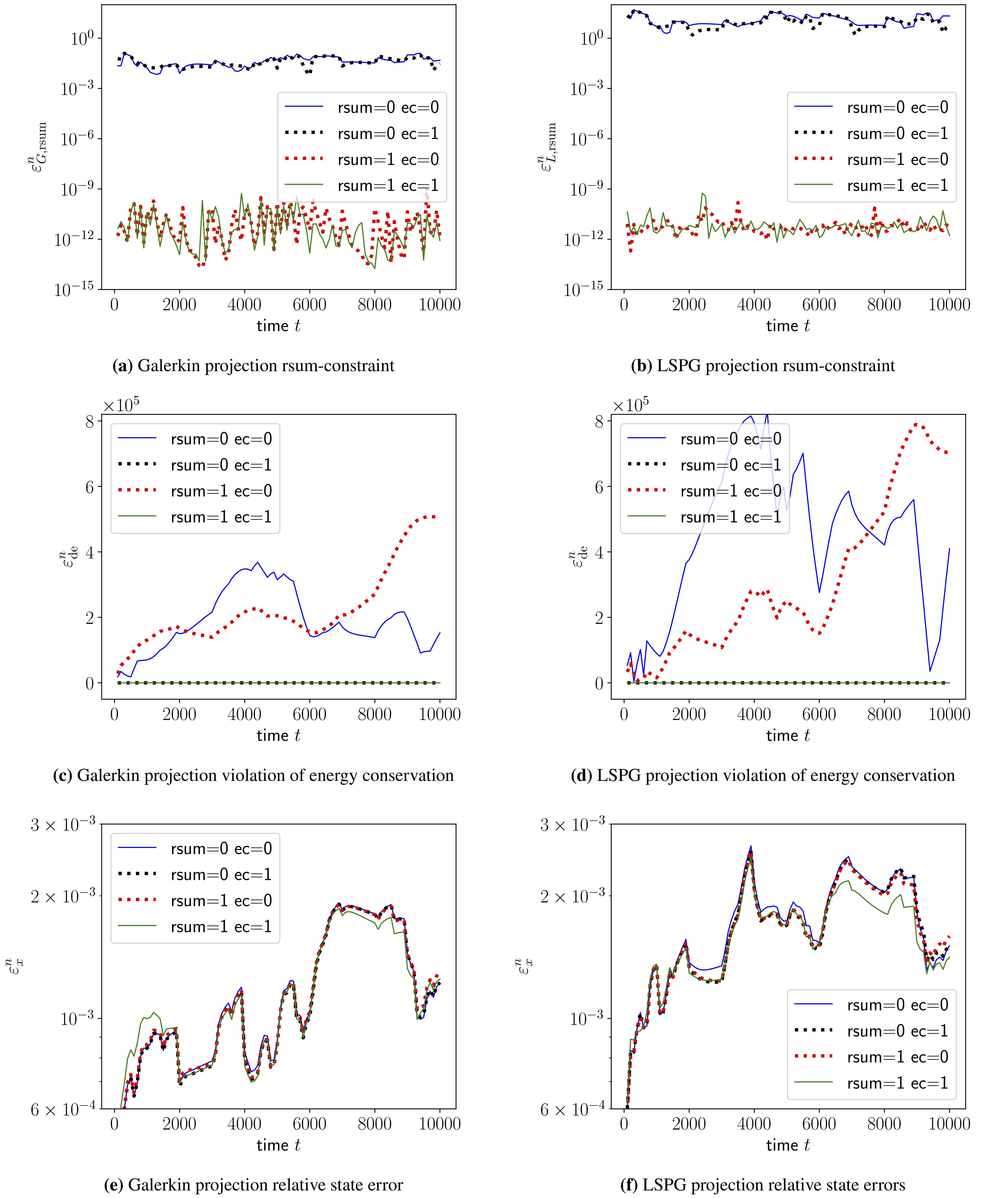} 
\caption{2D diffusion problem. Error metrics for the Galerkin projection (left column plots) and LSPG projection (right column plots) (10 degrees of freedom) evaluated at $\mv{\mu} = (10^6, -10^6, 10^{-4}, 0.2)$ with 100 time steps. rsum = 0/1 and ec = 0/1 denote the sum-of-residual-entries constraint and energy conservation constraint being \textit{not in use} / \textit{in use}.}
\label{fig:Galerkin projection diffusion problem}
\end{figure}

\begin{figure}[!htb]
\centering
\includegraphics[width=1.\linewidth]{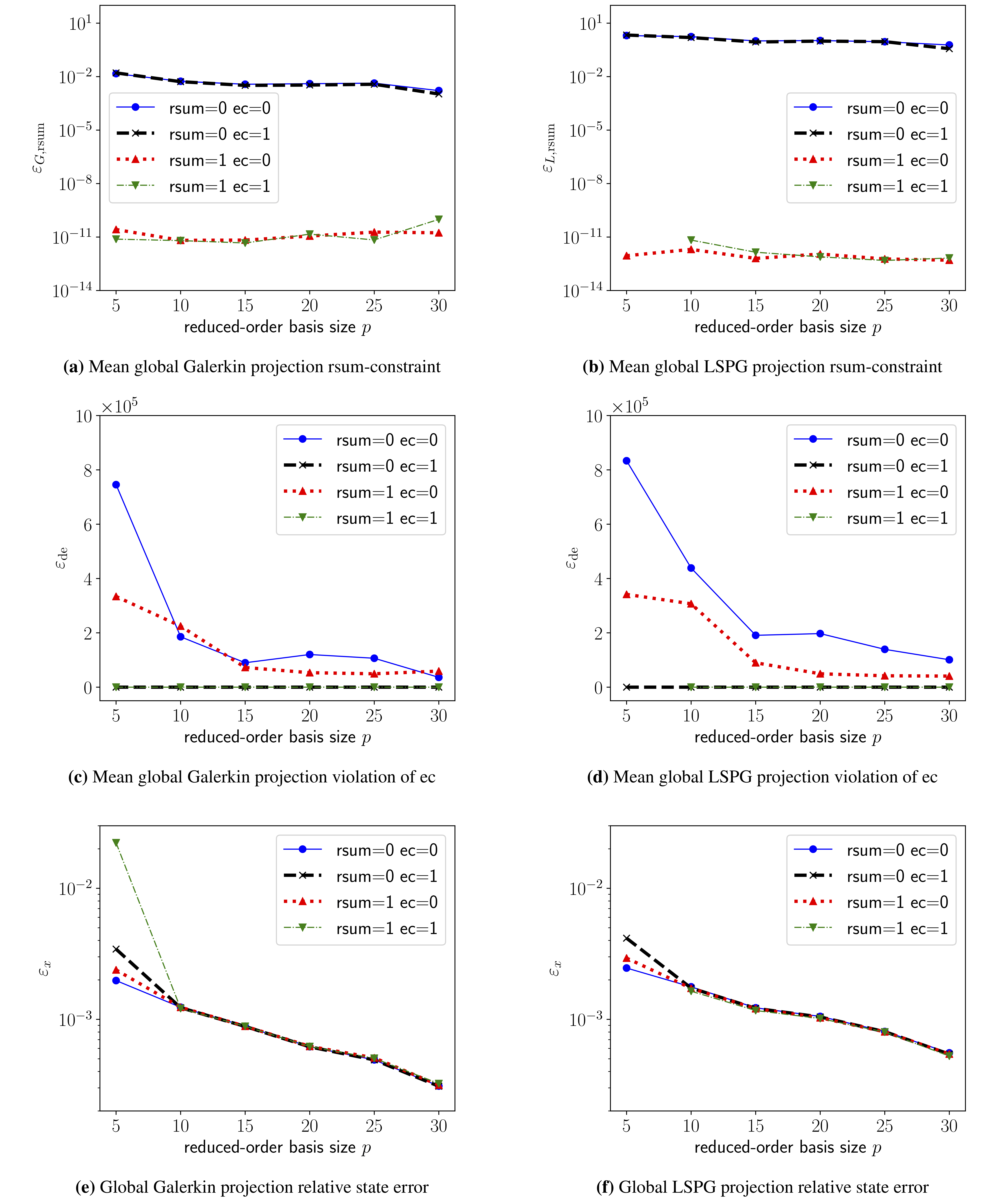} 
\caption{2D diffusion problem. Galerkin projection (left column) and LSPG projection (right column) global error metrics over variable basis size evaluated at $\mv{\mu} = (10^6, -10^6, 10^{-4}, 0.2)$ with 100 time steps . rsum = 0/1 and ec = 0/1 denote the sum-of-residual-entries constraint and energy conservation constraint being \textit{not in use} / \textit{in use}. The simulation rsum=1, ec=1, $p=5$ in case of the LSPG projection failed to converge and is therefore not depicted.}
\label{fig:Galerkin projection mean global error metrics}
\end{figure}

\begin{table}
\section{Problem-specific constraints}
\centering
\begin{tabular}[h]{ >{\centering}m{6cm} | >{\centering}m{4cm} | >{\centering}m{3cm} | c }
nomenclature & interpretation & constraint & reference \\
\hline 
\vbox{\begin{align*} \mv{\theta}&: \text{discretized  temperature field}
\\ \theta_l&: \text{lower temperature bound} \\ \theta_u&: \text{upper temperature bound} \end{align*}} & 
bounded temperature to physical range & 
\vbox{\begin{align*} \min(\mv{\theta}) - \theta_l \geq 0 \\ \theta_u - \max(\mv{\theta}) \geq 0 \end{align*}} & 
Ref.~\cite{Huang2019} \\
\hline
\vbox{\begin{align*} \bmv{C}&: \ \text{subdomain conservation  matrix} \\
\mv{r}&: \ \text{time-continuous \ residual} \\ \mv{r}^n&: \ \text{time-discrete  residual} \end{align*}} &
conservation in finite-volume models &
\vbox{\begin{align*} \bmv{C} \mv{r} &= \mv{0} \\ \bmv{C} \mv{r}^n &= \mv{0} \end{align*}} &
Ref.~\cite{Carlberg2018280} \\
\hline
\vbox{\begin{align*} C_{L, (\mathrm{target})}&: \mathrm{(target) \ lift \ coefficient} \\ C_{D, (\mathrm{target})}&: \mathrm{(target) \ drag \ coefficient} \end{align*}} &
fulfillment of aerodynamic constraints &
\vbox{\begin{align*} C_L - C_{L, \mathrm{target}} &= 0 \\ C_D - C_{D, \mathrm{target}} &= 0 \end{align*}} &
Ref.~\cite{Zimmermann2014255} \\
\hline
\vbox{\begin{align*} \mv{\lambda}: &\text{Lagrange  multipliers (in solid} \\
&\text{mechanics contact problem)} \end{align*}} & 
positivity of contact forces in solid mechanics contact problem &
$\mv{\lambda} \geq \mv{0}$ &
Ref.~\cite{balajewicz2016projection} \\
\hline
\vbox{\begin{align*} K&: \text{kinetic energy} \\ \mv{u}&: \text{velocity
field} \\ \nu&: \text{kinematic viscosity} \end{align*}}
&
kinetic energy balance in incompressible flows (under given modeling assumptions, see Ref.~\cite{sanderse2020non})
&
\vbox{\begin{equation*} \frac{d K}{d t} + \nu \norm{\nabla \mv{u}}^2 = 0 \end{equation*}} &
Ref.~\cite{sanderse2020non} \\
\hline
\vbox{\begin{align*} \hmv{x}_l&: \ \text{lower bounds on generalized
coordinates} \\ \hmv{x}_u&: \ \text{upper bounds on generalized coordinates} \end{align*}} &
bounded ROM solution by hyper-parameters estimated from FOM &
\vbox{\begin{align*} \hmv{x} - \hmv{x}_l &\geq \mv{0} \\ \hmv{x}_u - \hmv{x} &\geq \mv{0} \end{align*}} &
Ref.~\cite{Fick2018214} \\ 
\end{tabular}
\caption{Examples for constraints (with references) representable by the general constraint set \eqref{eq:kinematic equality constraint} - \eqref{eq:dynamic inequality constraint}. Note, that the constraint on bounded generalized coordinates (last row) is not directly representable by our general constraint set, which is defined independently from the ROM solution. We nevertheless include the constraint on bounded generalized coordinates, recalling that the generalized coordinates are expressible in terms of the FOM state $\hmv{x} = \mv{\Phi}^T (\mv{x} - \xref(\mv{\mu}))$.} 
\label{tab:literature overview}
\end{table}

\end{document}